\documentclass[a4paper,11pt]{article}
\pdfoutput=1 
\usepackage{jheppub} 
\usepackage{bm}
\usepackage{amssymb}
\usepackage[T1]{fontenc} 
\usepackage{graphicx}
\usepackage{subfigure}

\title{\boldmath Planck scale, Dirac delta function and  ultraviolet divergence}
\author[a,1]{Hua Zhang,\note{Corresponding author.}}
\author[b]{Mingshun Yuan}
\affiliation[a]{Shenzhen Forms Syntron Information Co.,Ltd.,\\ Shenzhen 518000, P.R. China}
\affiliation[b]{Fujian University of Technology,Mathematics and Physics Institute\\Fuzhou 350118, P.R. China}
\emailAdd{symtop@163.com}
\abstract{The Planck length is the minimum length which physical law do not fail. The Dirac delta function was created to deal with continuous range issue, and it is zero except for one point. Thus contradict the Planck length. Renormalization method  is the usual way to deal with divergence difficulties. The authors proposed a new way to solve the problem of ultraviolet divergence and this method is self-consistent with the Planck length. For this purpose a redefine function ${\delta}_P$ in position representation was introduced to handle with the canonical quantization. The function ${\delta}_P$ tends to the Dirac delta function if the Planck length goes to zero. By logical deduction the authors obtains new commutation/anticommutation relations and  new Feynman propagators which are convergent. This article will deduce the new Feynman propagators for the Klein-Gordon field, the Dirac field and the Maxwell field. Through the new Feynman propagators, we can eliminate  ultraviolet divergence. }

\begin{document} 
\maketitle
\flushbottom

\section{Introduction}
\label{sec:intro}

\subsection{History of renormalization}
Infinities problem arose when QFT constructed in the late 1920s or in the early 1930s. Many physicists devote to this hard work. If this problem can not be solved, QFT will be abandoned. After World War II, Sinitiro Tomonaga, Julian Schwinger, and Richard Feynman\cite{ref01,SchwingerJulian1948OQat,TomonagaSin-Itiro1948OIFR} found the solution independently, which is called renormalization theory of QED, and it is the most successful theory up to now. The successful calculation of the anomalous magnetic moment of the electron, announce the success of renormalization method. In 1971,Kenneth G. Wilson raised renormalization group\cite{WilsonKennethG.1971RGaS}, which changing cutoff $\Lambda$ of infinities will not impact the physical calculation results, for the interaction involve one or several momentum-dependent coupling constants. In 1972, Gerard 't Hooft and Martinus J.G. Veltman proved that non-abelian gauge fields  is renormalizable\cite{ref04}. A criterion\cite{ref02} for deciding which quantum field theories are renormalizable, that all infinities can be absorbed into a redefinition of a finite number of coupling constants and masses. The goal of this article is, infinities never arise in the procedure of calculation, we can achieve if not count infrared divergence.

\subsection{The Planck scale \&  the Dirac delta function}

In 1912, Max Planck given the  units of length\cite{ref05}, mass, time and temperature from the velocity $c$ of propagation of light in a vacuum, and the constant of gravitation $G$, and the Planck constant $h$. 

In 1999, Peter J. Mohr  and Barry N. Taylor  published an article list fundamental physical constants\cite{ref06}, with
\begin{align}
l_P&=({{\hbar}G/c^3})^{1/2}=1.6160(12)\times10^{-35}m \\
t_P&=l_P/c=({{\hbar}G/c^5})^{1/2}=5.3906(40)\times10^{-44}s
\end{align}

In 1927, in order to handle with continuous variables, Paul Dirac introduced a new generalized function\cite{ref07}  which named delta function ${\delta}(x)$. Delta function can be written as\cite{BjorkenJamesD1965Rqf}
\begin{equation}
\delta^3(\bm{x}-\bm{x}_0) =\lim_{\Delta{V}_i \to 0} \dfrac{\delta_{ij}}{\Delta{V}_i}
\end{equation}
Due to the limitation of Planck length, $\Delta{V}_i$ will not tend to zero. A new delta function can be defined as
\begin{equation}
\delta^3_P(\bm{x}-\bm{x}_0) = \dfrac{\delta_{ij}}{\Delta{V}_{Pi}}
\end{equation}
$\Delta{V}_{Pi}$ is the limit case of volume element $\Delta{V}_i$ in Planck length. In different coordinate system, the volume element $\Delta{V}_i$ has different form, thus ${\delta}_P$ has different form.\\
 Our deducing base on the below hypothesis.
\paragraph{i.}
   In position representation, the Dirac delta function (not include Kronecker delta function) in canonical commutation/anticommutation relations  must redefine as ${\delta}_P$ function.
\begin{equation}
{\delta} \to{\delta}_P
\end{equation}
\paragraph{ii.}
In canonical commutation/anticommutation relations, assume
\begin{equation}
{\int}d^3pe^{{\pm}ip.x}f(p) \to {\int}d^3pe^{{\pm}ip.x}{\Re}^3(\textbf{\textit{p}})f(p)
\end{equation}
Where $f(p)$ is the function of momentum $p$. For convenience, we call above item $\bm{i}.$ and item $\bm{ii}.$  renormalization transformations. On the action of renormalization transformations, ${\delta}_P$ function's Fourier expansion is called renormalization function, denote as ${\Re}^3(\textbf{\textit{p}})$. Renormalization function ${\Re}^3(\textbf{\textit{p}})$ in rectangular coordinate system represent as ${\Re}^3_{\bot}(\textbf{\textit{p}})$, in spherical coordinate system represent as ${\Re}^3_s(\textbf{\textit{p}})$. Below paragraphs will given the expressions of   ${\Re}^3(\textbf{\textit{p}})$ and ${\delta}_P$.

In this paper, the renormalization functions are introduced, and the explicit expression of the renormalization functions are given in rectangular and cylindrical coordinates. In Klein-Gordon field, Dirac field and electromagnetic field,  new standard commutation, anticommutation relations and  new Feynman propagators are proposed. The authors proved that the loop diagrams of vacuum polarization and electron self-energy are convergent in large momentum when introduced renormalization functions, and can be extend to arbitrary order case(for rectangular and cylindrical systems). In the appendix, the authors calculate the masses of octet baryons use renormalization functions(in rectangular  and cylindrical system).

\section{The renormalization function }
\subsection{Rectangular coordinate system}

In the Planck length limit, the particles have determinate place ${\textbf{\textit{x}}}{\in}({\textbf{\textit{x}}}_0-l_P/2,{\textbf{\textit{x}}}_0+l_P/2)$, uncertainty ${\Delta}{\textbf{\textit{x}}}=l^3_P=\Delta{V}$. Wave function
\begin{equation}
{\Psi}_{x_0}({\textbf{\textit{x}}})={\delta}^3_P({\textbf{\textit{x}}}-{\textbf{\textit{x}}}_0)
\end{equation}
 In this article, if a symbol is added a subscript $P$ or a superscript $P$, it represents the corresponding symbol under the renormalization transformations.
\begin{equation}
 {\delta}^3_P({\textbf{\textit{x}}}-{\textbf{\textit{x}}}_0)= \begin{cases}
1/l^3_P, {\textbf{\textit{x}}}{\in}({\textbf{\textit{x}}}_0-l_P/2,{\textbf{\textit{x}}}_0+l_P/2)\\
0,\qquad {\textbf{\textit{x}}}{\not\in}({\textbf{\textit{x}}}_0-l_P/2,{\textbf{\textit{x}}}_0+l_P/2) \\ 
\end{cases}
\end{equation}
For ${\delta}^3_P({\textbf{\textit{x}}}-{\textbf{\textit{x}}}_0)$   its Fourier expansion 
\begin{equation}
{\Psi}_{x_0}({\textbf{\textit{p}}})=\dfrac{1}{\sqrt{({2\pi})^3} }e^{-ip.x_0}{\Re}^3_{\bot}(\textbf{\textit{p}})
\end{equation}
With
\begin{align}
{\Re}^3_{\bot}(\textbf{\textit{p}})&=[{\sin\dfrac{l_Pp_x}{2\hbar }}/{\dfrac{l_Pp_x}{2\hbar}}][{\sin\dfrac{l_Pp_y}{2\hbar }}/{\dfrac{l_Pp_y}{2\hbar}}][{\sin\dfrac{l_Pp_z}{2\hbar }}/{\dfrac{l_Pp_z}{2\hbar}}]
\end{align}
Compare with Fourier expansion of the delta funtion ${\delta}^3({\textbf{\textit{x}}}-{\textbf{\textit{x}}}_0)$, we got a new item ${\Re}^3_{\bot}(\textbf{\textit{p}})$, When $l_P$ is infinitesimal,
\begin{align}
\lim_{l_P \to 0}{\delta}^3_P({\textbf{\textit{x}}}-{\textbf{\textit{x}}}_0)&={\delta}^3({\textbf{\textit{x}}}-{\textbf{\textit{x}}}_0) \\
\lim_{l_P \to 0}\dfrac{1}{\sqrt{({2\pi})^3} }e^{-ip.x_0}{\Re}^3_{\bot}(\textbf{\textit{p}})&=\dfrac{1}{\sqrt{({2\pi})^3} }e^{-ip.x_0}
\end{align}
We call ${\Re}^3_{\bot}(\textbf{\textit{p}})$  as renormalization function in rectangular coordinate system, or call ${\Re}^3_{\bot}(\textbf{\textit{p}})$  as momentum distribution wave function in rectangular coordinate system. Compare the renormalization function with Fraunhofer moment aperture diffraction intensity distribution formula, they have similar expression.

\begin{figure}[tbp]
\centering 
\includegraphics[width=1\textwidth]{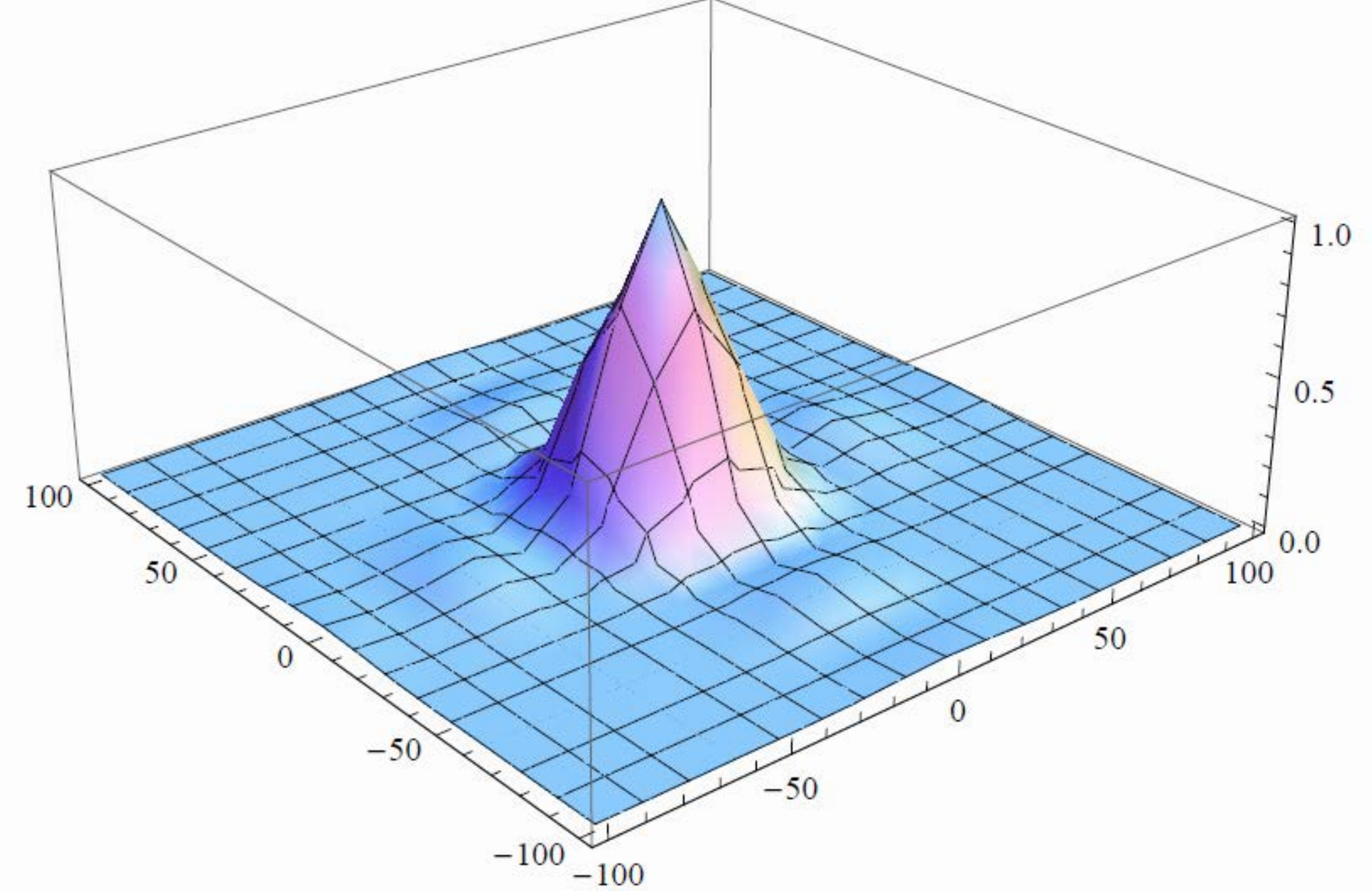}
\hfill
\caption{\label{fig:i}Renormalization function ${\Re}^2_{\bot}(p)=[{\sin\dfrac{l_Pp_x}{2\hbar }}/{\dfrac{l_Pp_x}{2\hbar}}][{\sin\dfrac{l_Pp_y}{2\hbar }}/{\dfrac{l_Pp_y}{2\hbar}}]$ }, where x-axis and y-axis represent $p_x$ and $p_y$, z-axis represent ${\Re}^2_{\bot}(p)$.
\end{figure}

\subsection{Cylindrical coordinate system}
In the Planck length limit  of cylindrical coordinate system $\Delta{V} = \frac{{\pi}l^3_P}{4}$, the delta function 
\begin{equation}
 {\delta}^3_{Pcylin}(r,\theta,z)= \begin{cases}
    \frac{4}{{\pi}l^3_P} , r<{l_P}/2, \left|z\right|<{l_P}/2 \\

0,\quad  r\geq{l_P}/2 , \left|z\right|\geq{l_P}/2\\ 
\end{cases}
\end{equation}
Corresponding Fourier expansion
\begin{equation}
{\Psi}_{x_0}({\textbf{\textit{p}}})=\dfrac{1}{\sqrt{({2\pi})^3} }e^{-ip.x_0}{\Re}^3_{cylin}(\textbf{\textit{p}})
\end{equation}
Where
\begin{align}
{\Re}^3_{cylin}(\textbf{\textit{p}}) &=  \frac{4}{{\pi}l^3_P} \int_{-l_P/2}^{l_P/2} \int_{0}^{2\pi} \int_{0}^{l_P/2} r 
{e^{i[(p_x\cos\theta+p_y\sin\theta)r+p_z z]/\hbar} }  dr d\theta dz \\
 &=  \sideset{_0}{}{\mathop{F_1}}(2;-\frac{l_P^2(p_x^2+p_y^2)}{16\hbar^2})[{\sin\dfrac{l_Pp_z}{2\hbar }}/{\dfrac{l_Pp_z}{2\hbar}}]
\end{align}
Where $\sideset{_0}{}{\mathop{F_1}}(a;z)$ is the generalized hypergeometric series (also see Figure 2./Figure 3./Figure 4., and see \cite{BaileyWilfridNorman1935Ghs,BieberbachLudwig1953TdGD}), satisfy
\begin{align}
 \sideset{_0}{}{\mathop{F_1}}(2;-\frac{l_P^2(p_x^2+p_y^2)}{16\hbar^2}) &= \sum_{n=0}^\infty \frac{(-1)^n}{n!(n+1)!}[\frac{l_P^2(p_x^2+p_y^2)}{16\hbar^2}]^n\\
\lim_{l_P \to 0}  \sideset{_0}{}{\mathop{F_1}}(2;-\frac{l_P^2(p_x^2+p_y^2)}{16\hbar^2}) &= 1
\end{align}
We have
\begin{align}
\lim_{l_P \to 0}{\Re}^3_{cylin}(\textbf{\textit{p}})&=1
\end{align}

\begin{figure}[tbp]
\centering 
\includegraphics[width=1\textwidth]{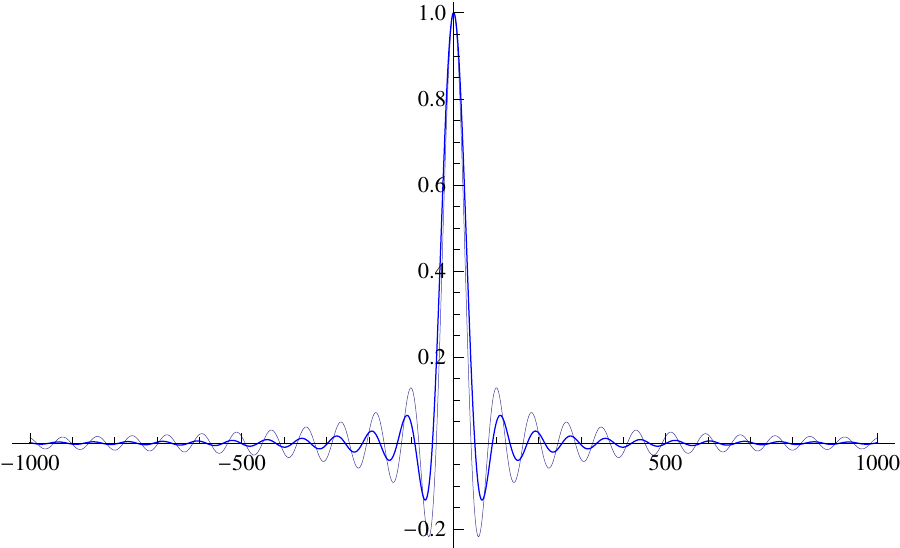}
\hfill
\caption{\label{fig:i}The thick line  is the curve of renormalization function ${\Re}^3_{cylin}(\bm{p})|_{p_y=0,p_z=0}=\sideset{_0}{}{\mathop{F_1}}(2;-\frac{l_P^2 p_x^2}{16\hbar^2})$ }; The thin line is the curve of renormalization function  ${\sin\dfrac{l_Pp_x}{2\hbar }}/{\dfrac{l_Pp_x}{2\hbar}}$; x-axis represents $p_x$.
\end{figure}

\begin{figure}[tbp]
\centering 
\includegraphics[width=1\textwidth]{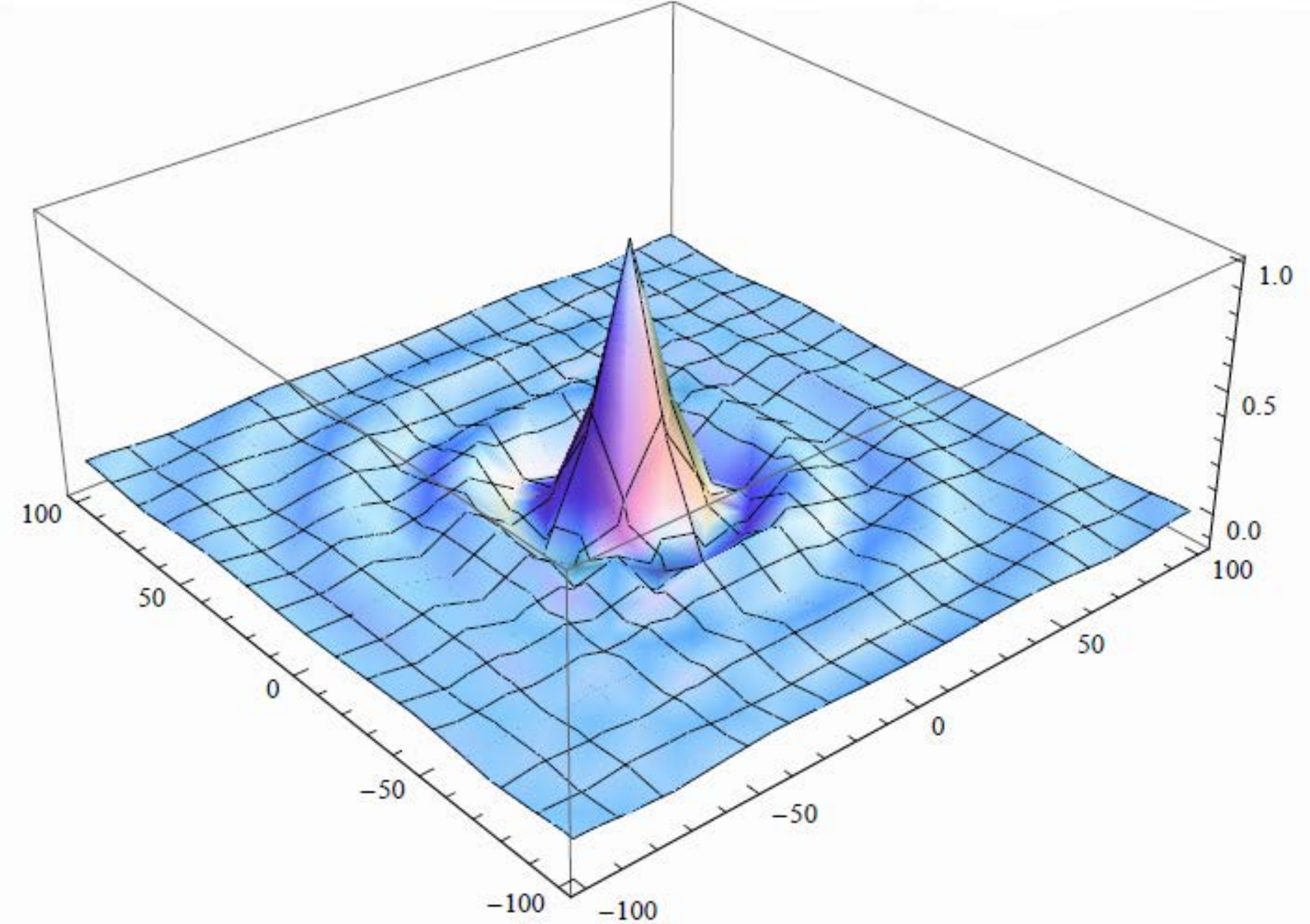}
\hfill
\caption{\label{fig:i}Renormalization function ${\Re}^3_{cylin}(\bm{p})|_{p_z=0}=\sideset{_0}{}{\mathop{F_1}}(2;-\frac{1}{16\hbar^2}l_P^2(p_x^2+p_y^2))$ }, where x-axis and y-axis represent $p_x$ and $p_y$, z-axis represent ${\Re}^3_{cylin}(\bm{p})|_{p_z=0}$.
\end{figure}

\begin{figure}[htbp]
\centering
\subfigure[]{
\begin{minipage}[t]{0.5\linewidth}
\centering
\includegraphics[width=1\textwidth]{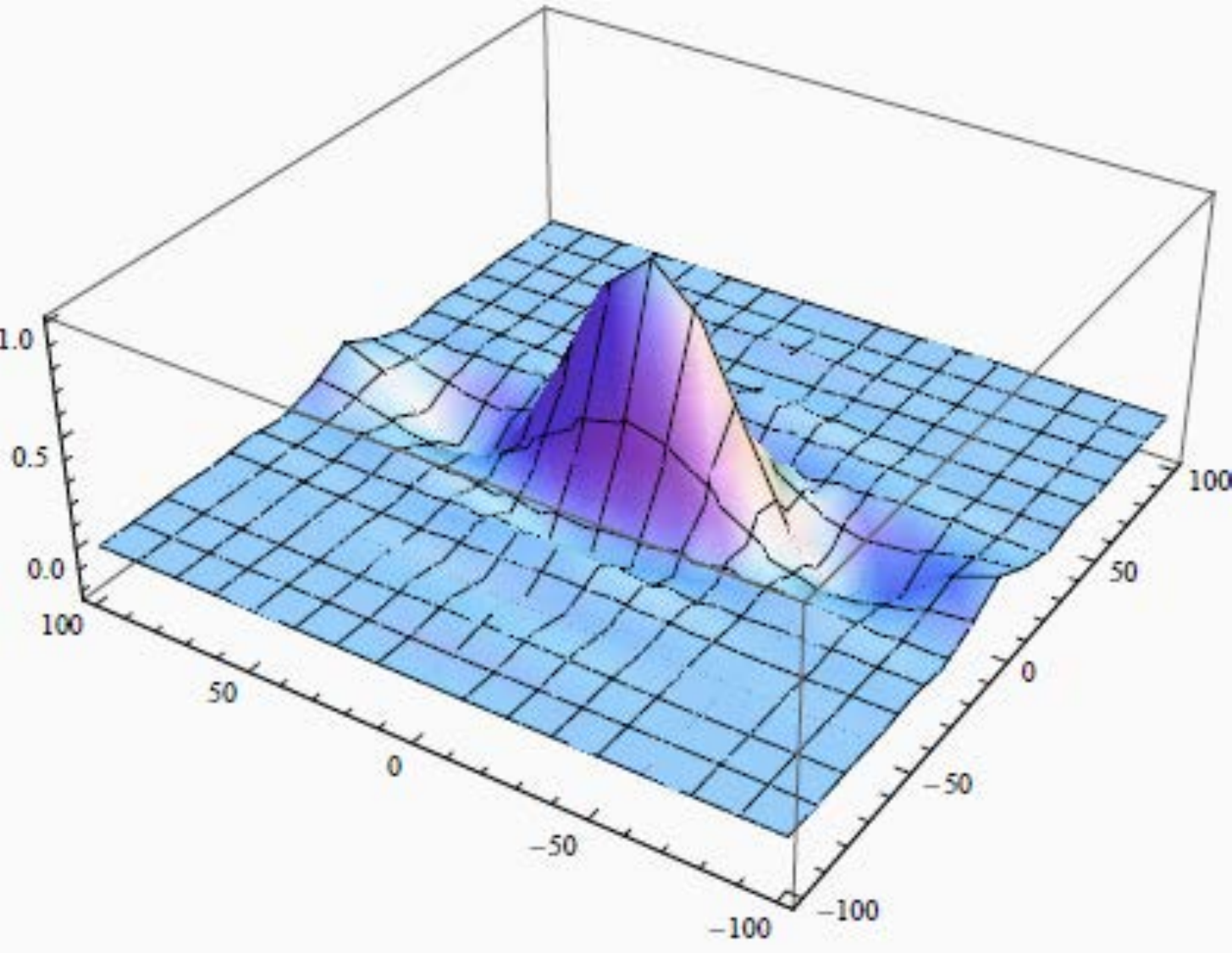}
\end{minipage}%
}%
\subfigure[]{
\begin{minipage}[t]{0.5\linewidth}
\centering
\includegraphics[width=1\textwidth]{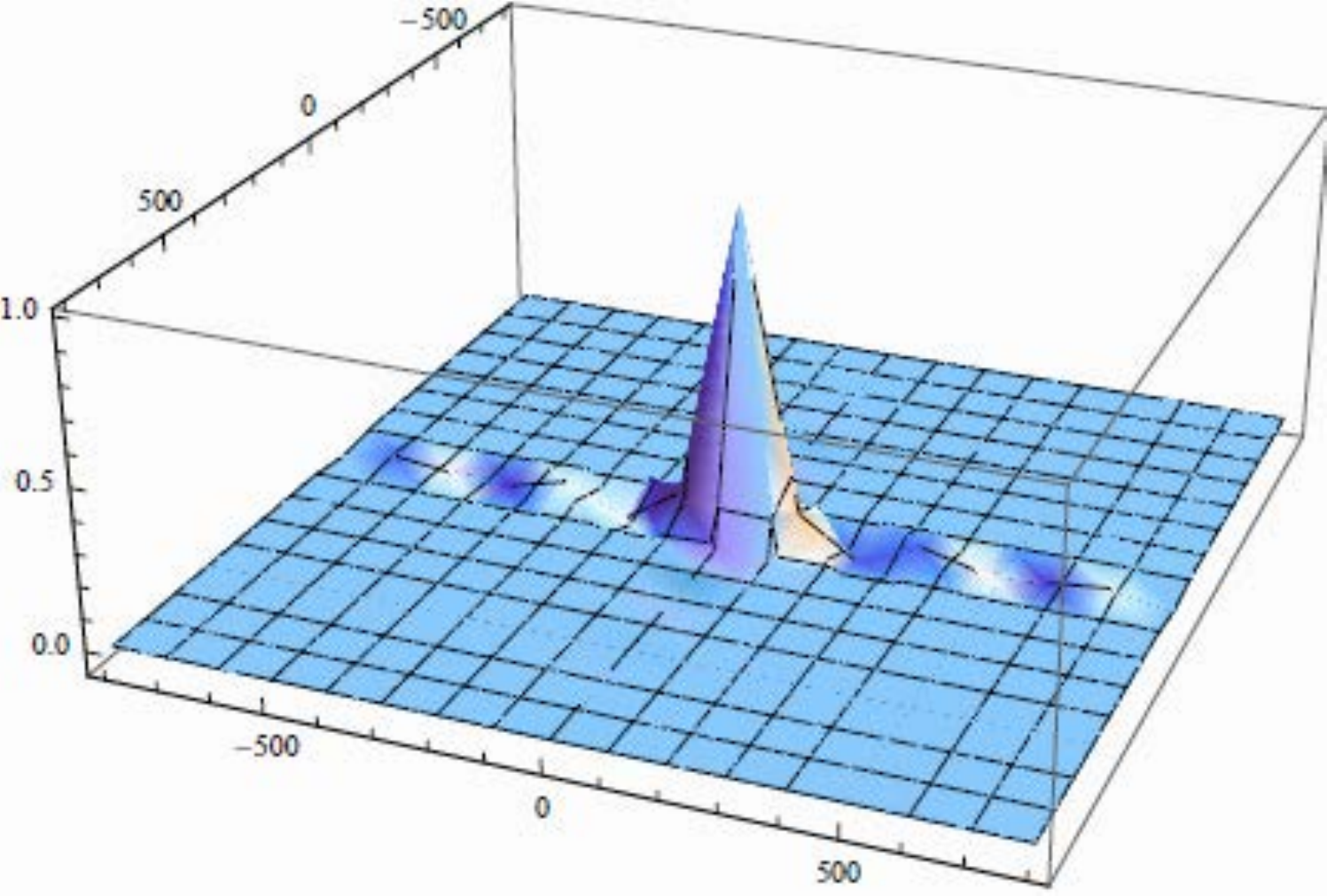}
\end{minipage}%
}%
\centering
\caption{ Renormalization function ${\Re}^2_{cylin}(\bm{p})=\sideset{_0}{}{\mathop{F_1}}(2;-\frac{1}{16\hbar^2}l_P^2 p_x^2)[{\sin\dfrac{l_Pp_y}{2\hbar }}/{\dfrac{l_Pp_y}{2\hbar}}]$}
\end{figure}

\subsection{Spherical coordinate system}
In the Planck length limit  of spherical coordinate system $\Delta{V} = \frac{{\pi}l^3_P}{6}$, the delta function 
\begin{equation}
 {\delta}^3_{Ps}(r,\theta,\varphi)= \begin{cases}
    \frac{6}{{\pi}l^3_P} , r<{l_P}/2  \\

0,\quad  r\geq{l_P}/2 \\ 
\end{cases}
\end{equation}
Corresponding Fourier expansion
\begin{equation}
{\Psi}_{x_0}({\textbf{\textit{p}}})=\dfrac{1}{\sqrt{({2\pi})^3} }e^{-ip.x_0}{\Re}^3_s(\textbf{\textit{p}})
\end{equation}
Where
\begin{align}
{\Re}^3_s(\textbf{\textit{p}}) &= \frac{6}{{\pi}l^3_P} \int_{0}^{2\pi} \int_{0}^{\pi} \int_{0}^{l_P/2} r^2\sin\varphi 
 {e^{i(p_x\sin\varphi\cos\theta+p_y\sin\varphi\sin\theta+p_z\cos\varphi)r/\hbar} }  dr  d\varphi  d\theta \\
&= \frac{12}{l^3_P} \int_{0}^{\pi}  \int_{0}^{l_P/2} r^2\sin\varphi 
 {e^{i(p_z\cos\varphi)r/\hbar} } J_0(\frac{r\sin\varphi\sqrt{p^2_x+p^2_y}}{\hbar}) dr  d\varphi 
\end{align} 
with the first kind of Bessel function of order 0(see \cite{WatsonG.N1944Atot})
\begin{align}
 J_0(\frac{r\sin\varphi\sqrt{p^2_x+p^2_y}}{\hbar}) = \frac{1}{2\pi}\int_{0}^{2\pi}e^{i(p_x\sin\varphi\cos\theta+p_y\sin\varphi\sin\theta)r/\hbar}   d\theta 
\end{align}
We obtain
\begin{align}
\lim_{l_P \to 0}{\delta}^3_{Ps}(r,\theta,\varphi)&={\delta}^3({\textbf{\textit{x}}}-{\textbf{\textit{x}}}_0)\\
\lim_{l_P \to 0}\dfrac{1}{\sqrt{({2\pi})^3} }e^{-ip.x_0}{\Re}^3_s(\textbf{\textit{p}})&=\dfrac{1}{\sqrt{({2\pi})^3} }e^{-ip.x_0}
\end{align}
${\Re}^3_s(\textbf{\textit{p}})$   denote as renormalization function in spherical coordinate system. Since the explicit expression of  ${\Re}^3_s(\textbf{\textit{p}})$  cannot be obtained, this paper will not discuss it in detail.

\section{Quantization and  propagator }
\subsection{The Klein-Gordon field}
The usual canonical commutation relations may be written\cite{BjorkenJamesD1965Rqf,Itzykson011}
\begin{align}
[\phi_P(\textbf{\textit{x}},t),\pi_P(\textbf{\textit{y}},t)] &= i{\delta}^3_P(\textbf{\textit{x}}-{\textbf{\textit{y}}}) \\
[\phi_P(\textbf{\textit{x}},t),\phi_P(\textbf{\textit{y}},t)] &=0 \\
[\pi_P(\textbf{\textit{x}},t),\pi_P(\textbf{\textit{y}},t)] &=0
\end{align}
Here with a ${\delta}^3_P(\textbf{\textit{x}}-{\textbf{\textit{y}}})$  in place of a delta function  ${\delta}^3(\textbf{\textit{x}}-{\textbf{\textit{y}}})$.
And 
\begin{equation}
\phi_P(\textbf{\textit{x}},t)= \int \dfrac{d^3k}{\sqrt{({2\pi})^3 {2\omega_k}} }[a_P(\textbf{\textit{k}})e^{-ik.x}+a_P^\dagger(\textbf{\textit{k}})e^{ik.x}]
\end{equation}
 The commutation relations of annihilate operator and create operator under the renormalization transformation, will change to
\begin{align}
[a_P(\textbf{\textit{k}}),a^\dagger_P(\textbf{\textit{k}}^\prime)] &=  {\Re}^3(\textbf{\textit{k}}^\prime) {\delta}^3(\textbf{\textit{k}}-{\textbf{\textit{k}}}^\prime)  \\
[a_P(\textbf{\textit{k}}),a_P(\textbf{\textit{k}}^\prime)] &= 0  \\
[a^\dagger_P(\textbf{\textit{k}}),a^\dagger_P(\textbf{\textit{k}}^\prime)] &=0
\end{align}
Annihilate operator and create operator can be written
\begin{align}
a_P(\bm{k}) &= \sqrt{{\Re}^3(\bm{k})}a(\bm{k}) \\
a^\dagger_P(\bm{k}) &= \sqrt{{\Re}^3(\bm{k})}^\dagger a^\dagger(\bm{k})
\end{align}
Now we can calculate the Feynman propagator from above commutation relations. For the Klein-Gordon field the Feynman propagator may be written(see Figure 5(a))
\begin{align}
[i\Delta_F(x,y)]_P &= \int \dfrac{d^3k}{({2\pi})^3 2\omega_k}[ e^{-ik.(x-y)} {\Re}^3(\textbf{\textit{k}})\Theta(x_0-y_0)
+ e^{ik.(x-y)} {\Re}^3(\textbf{\textit{k}})\Theta(y_0-x_0) ]{} \nonumber \\
&,\omega_k=\sqrt{\textbf{\textit{k}}^2+m^2}  
\end{align} 
Where
\[ \Theta(x_0-y_0)= \begin{cases}
1,\quad x_0>y_0 \\
0,\quad x_0<y_0
\end{cases} \]
The Feynman propagator can be written in a unified form
\begin{align}
[i\Delta_F(x,y)]_P &= i\int \dfrac{d^4k}{({2\pi})^4} e^{-ik.(x-y)}\dfrac{ {\Re}^3(\textbf{\textit{k}}) }{k^2-m^2+i\epsilon}
\end{align} 

\subsection{The Dirac field}
After the renormalization transformation the usual canonical anticommutation relations may be written\cite{BjorkenJamesD1965Rqf}
\begin{align}
[\pi^P_\alpha(\textbf{\textit{x}},t),\psi^P_\beta(\textbf{\textit{y}},t)]_+ &= i{\delta}^3_P(\textbf{\textit{x}}-{\textbf{\textit{y}}}) \delta_{\alpha\beta}\\
[\pi^P_\alpha(\textbf{\textit{x}},t),\pi^P_\beta(\textbf{\textit{y}},t)]_+ &= 0\\
[\psi^P_\alpha(\textbf{\textit{x}},t),\psi^P_\beta(\textbf{\textit{y}},t)]_+ &= 0 
\end{align}
Here with a ${\delta}^3_P(\textbf{\textit{x}}-{\textbf{\textit{y}}})$  in place of a delta function  ${\delta}^3(\textbf{\textit{x}}-{\textbf{\textit{y}}})$.
And 
\begin{align}
\psi_P(\textbf{\textit{x}},t)&=\sum_{\pm s} \int d^3p\sqrt{\dfrac{m}{({2\pi})^3 {\omega_p}} }[b_P(\textbf{\textit{p}},s)u(\textbf{\textit{p}},s)e^{-ip.x} 
+d_P^\dagger(\textbf{\textit{p}},s)v(\textbf{\textit{p}},s)e^{ip.x}]
\end{align}
We can obtain below results from  the  above canonical anticommutation relations
\begin{align}
[b_P(\textbf{\textit{p}},s),b^\dagger_P(\textbf{\textit{p}}^\prime,s^\prime)]_+ &=  {\Re}^3(\textbf{\textit{p}})   \delta_{ss^\prime} {\delta}^3(\textbf{\textit{p}}-{\textbf{\textit{p}}}^\prime)  \\
[d_P(\textbf{\textit{p}},s),d^\dagger_P(\textbf{\textit{p}}^\prime,s^\prime)]_+ &=  {\Re}^3(\textbf{\textit{p}})   \delta_{ss^\prime} {\delta}^3(\textbf{\textit{p}}-{\textbf{\textit{p}}}^\prime)  \\
[b_P,b_P]_+ = [d_P,d_P]_+ &=[b_P,d_P]_+ = 0 \\
[b^\dagger_P,b^\dagger_P]_+ = [d^\dagger_P,d^\dagger_P]_+ &=[b^\dagger_P,d^\dagger_P]_+ = 0\\
[b_P,d^\dagger_P]_+ = [b^\dagger_P,d_P]_+ &= 0
\end{align}
Annihilate operator and create operator can be written
\begin{align}
b_P(\bm{k}) &= \sqrt{{\Re}^3(\bm{k})}b(\bm{k}) \\
b^\dagger_P(\bm{k}) &= \sqrt{{\Re}^3(\bm{k})}^\dagger b^\dagger(\bm{k})\\
d_P(\bm{k}) &= \sqrt{{\Re}^3(\bm{k})}d(\bm{k}) \\
d^\dagger_P(\bm{k}) &= \sqrt{{\Re}^3(\bm{k})}^\dagger d^\dagger(\bm{k})
\end{align}
the Feynman propagator(see Figure 5(b))
\begin{align}
[iS_F(x,y)]_P &= \int \dfrac{d^3p}{({2\pi})^3 2\omega_p}[ e^{-ip.(x-y)} {\Re}^3(\textbf{\textit{p}})(m+{p\mkern-8mu/})
\Theta(x_0-y_0)+ e^{ip.(x-y)} {\Re}^3(\textbf{\textit{p}})(m-{p\mkern-8mu/}) \nonumber \\
& \times\Theta(y_0-x_0)], \omega^2_p= \textbf{\textit{p}}^2+m^2 
\end{align} 
it can be  written in a unified form
\begin{align}
[iS_F(x,y)]_P &=i\int \dfrac{d^4p}{({2\pi})^4} e^{-ip.(x-y)}   \dfrac{ {\Re}^3(\textbf{\textit{p}})}{{p\mkern-8mu/} - m+i\epsilon} 
\end{align}

\subsection{The Maxwell field}
In this section the canonical quantization may use radiation gauge. Under the renormalization transformations the equal-time commutators are\cite{BjorkenJamesD1965Rqf}
\begin{align}
[A^P_i(\textbf{\textit{x}},t),\pi^j_P(\textbf{\textit{y}},t)] &= i\delta_{ij} \delta^3_P(\textbf{\textit{x}}-\textbf{\textit{y}})-i\int \dfrac{d^3k}{({2\pi})^3}   
 e^{i\textbf{\textit{k}}.(\textbf{\textit{x}}-\textbf{\textit{y}})}{\Re}^3(\textbf{\textit{k}})(\dfrac{k_ik_j}{\textbf{\textit{k}}^2 })  \\
[A^P_i(\textbf{\textit{x}},t),A^P_j(\textbf{\textit{y}},t)] &= [\pi^i_P(\textbf{\textit{x}},t),\pi^j_P(\textbf{\textit{y}},t)] = 0
\end{align}
For the Maxwell field 
\begin{align}
A^P_\mu(\textbf{\textit{x}},t)= {}&\int \dfrac{d^3k}{\sqrt{({2\pi})^3 {2\omega_k}} }\sum_{\lambda=1}^2 \epsilon_\mu(k,\lambda)[a_{P}(\textbf{\textit{k}},\lambda)e^{-ik.x}+a_{P}^\dagger(\textbf{\textit{k}},\lambda)e^{ik.x}]
\end{align}
where 
\begin{align}
\epsilon_\mu(k,\lambda) = (0,\bm{\epsilon}(k,\lambda))
\end{align}
with $\epsilon(k,\lambda)$  is polarization vector, satisfy
\begin{align}
\textbf{\textit{k}}.\bm{\epsilon}(k,\lambda) = 0,\lambda &=1,2 \\
\bm{\epsilon}(k,\lambda).\bm{\epsilon}(k,\lambda^\prime)  &=\delta_{\lambda\lambda^\prime}\\
\bm{\epsilon}(-k,1) =-\bm{\epsilon}(k,1),\bm{\epsilon}(-k,2)  &=+\bm{\epsilon}(k,2)
\end{align}
We can obtain below results for annihilate operator and create operator commutation relations
\begin{align}
[a_{P}(\textbf{\textit{k}},\lambda),a^\dagger_{P}(\textbf{\textit{k}}^\prime,\lambda^\prime)] &= [({\Re}^3(\textbf{\textit{k}})+ {\Re}^3(-\textbf{\textit{k}}) )/2]{\delta}^3(\textbf{\textit{k}}-{\textbf{\textit{k}}}^\prime)\delta_{\lambda\lambda^\prime}  \\
[a_{P}(\textbf{\textit{k}},\lambda),a_{P}(\textbf{\textit{k}}^\prime,\lambda^\prime)] &= [a^\dagger_{P}(\textbf{\textit{k}},\lambda),a^\dagger_{P}(\textbf{\textit{k}}^\prime,\lambda^\prime)] = 0
\end{align}
Annihilate operator and create operator can be written
\begin{align}
a_P(\bm{k},\lambda) &= \sqrt{\dfrac{{\Re}^3(\bm{k})+ {\Re}^3(-\bm{k})}{2}}a(\bm{k},\lambda) \\
a^\dagger_P(\bm{k},\lambda) &= \sqrt{\dfrac{{\Re}^3(\bm{k})+ {\Re}^3(-\bm{k}) }{2}}^\dagger a^\dagger(\bm{k},\lambda)
\end{align}
 The photon propagator(see Figure 5(c))
\begin{align}
[iD_{\mu\nu}(x,y)]_P &= \int \dfrac{d^3k}{({2\pi})^3 2\omega_k}\sum_{\lambda=1}^2 \epsilon_\mu(k,\lambda) \epsilon_\nu(k,\lambda)  [({\Re}^3(\textbf{\textit{k}})+ {\Re}^3(-\textbf{\textit{k}}) )/2]\times[e^{-ik.(x-y)}\Theta(x_0-y_0){}\nonumber\\
&+e^{ik.(x-y)}\times\Theta(y_0-x_0) ]
\end{align} 
with $\omega_k=\sqrt{\textbf{\textit{k}}^2+m^2_\gamma}=\sqrt{\textbf{\textit{k}}^2}$.
When we set
\begin{align}
\eta^\mu = (1,0,0,0),\hat{k}^\mu = \dfrac{k^\mu-(k.\eta)\eta^\mu}{\sqrt{(k.\eta)^2-k^2}}
\end{align}
We may obtain
\begin{align}
\sum_{\lambda=1}^2 \epsilon_\mu(k,\lambda) \epsilon_\nu(k,\lambda) &=-g_{\mu\nu}-\dfrac{k_\mu k_\nu}{(k.\eta)^2-k^2}+\dfrac{(k.\eta)(k_\mu \eta_\nu+\eta_\mu k_\nu)}{(k.\eta)^2-k^2}-\dfrac{k^2 \eta_\mu \eta_\nu}{(k.\eta)^2-k^2}
\end{align}
With $g_{\mu\nu}$  is metric in spacetime.Photon propagator can be written in a unified form
\begin{align}
[iD_{\mu\nu}(x,y)]_P &= i\int \dfrac{d^4k}{({2\pi})^4}\sum_{\lambda=1}^2 \epsilon_\mu(k,\lambda) \epsilon_\nu(k,\lambda)  [({\Re}^3(\textbf{\textit{k}})
+ {\Re}^3(-\textbf{\textit{k}}) )/2]\times \dfrac{e^{-ik.(x-y)} }{k^2+i\epsilon}
\end{align} 

\begin{figure}[htbp]
\centering
\subfigure[Klein-Gordan field]{
\begin{minipage}[t]{0.3\linewidth}
\centering
\includegraphics[width=1\textwidth]{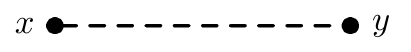}
\end{minipage}%
}%
\subfigure[Dirac field]{
\begin{minipage}[t]{0.3\linewidth}
\centering
\includegraphics[width=1\textwidth]{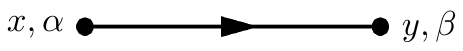}
\end{minipage}%
}%
\centering
\subfigure[Maxwell field]{
\begin{minipage}[t]{0.3\linewidth}
\centering
\includegraphics[width=1\textwidth]{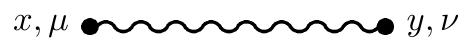}
\end{minipage}%
}%
\centering
\caption{Propergators for the Klein-Gordan, Dirac, Maxwell fields}
\end{figure}
\section{Feynman loop diagram}
This section will discuss Feynman loop diagrams, the corresponding renormalization function is in rectangular coordinate system or cylindrical coordinate system. The proof that the Feynman integral of loop diagram is convergent refer to appendix.
\subsection{Vacuum polarization}
The Feynman integral of Figure 6(a).
\begin{align}
\Pi^{\mu\nu}(k) = \int \dfrac{d^4q}{{(2\pi)}^4}(-1)tr[(-ie\gamma^\mu)\dfrac{i{\Re}^3(\bm{q}-\bm{k})}{{q\mkern-8mu/}-{k\mkern-8mu/} - m+i\epsilon}(-ie\gamma^\nu)\dfrac{i{\Re}^3(\bm{q})}{{q\mkern-8mu/} - m+i\epsilon}]
\end{align} 
Above equation can be changed to 
\begin{align}
\Pi^{\mu\nu}(k) = \int \dfrac{d^3q}{16\pi^2}tr[(-ie\gamma^\mu)\dfrac{i({q\mkern-8mu/}-{k\mkern-8mu/}+m){\Re}^3(\bm{q}-\bm{k})}{\omega_{q-k}}(-ie\gamma^\nu)\dfrac{i({q\mkern-8mu/}+m){\Re}^3(\bm{q})}{\omega_q}]
\end{align} 
with
\begin{align}
\omega_{q-k} &=\sqrt{(\bm{q}-\bm{k})^2+m^2}  \\
\omega_q &=\sqrt{\bm{q}^2+m^2} 
\end{align} 
Remove the items of equation (4.1)  whose trace are zero 
\begin{align}
\Pi^{\mu\nu}(k) &= \int \dfrac{d^3q}{16\pi^2}tr[(-ie\gamma^\mu)\dfrac{i({q\mkern-8mu/}-{k\mkern-8mu/}){\Re}^3(\bm{q}-\bm{k})}{\omega_{q-k}}(-ie\gamma^\nu)\dfrac{i{q\mkern-8mu/}{\Re}^3(\bm{q})}{\omega_q}] \nonumber\\
&+m^2\int \dfrac{d^3q}{16\pi^2}tr[(-ie\gamma^\mu)\dfrac{i{\Re}^3(\bm{q}-\bm{k})}{\omega_{q-k}}(-ie\gamma^\nu)\dfrac{i{\Re}^3(\bm{q})}{\omega_q}]\\
 &= e^2 tr[\gamma^\mu\gamma^\rho\gamma^\nu\gamma^\lambda]\int \dfrac{d^3q}{16\pi^2} \dfrac{(q_\rho-k_\rho)q_\lambda{\Re}^3(\bm{q}-\bm{k})}{\omega_{q-k}}\dfrac{{\Re}^3(\bm{q})}{\omega_q} \\
&+m^2e^2 tr[\gamma^\mu\gamma^\nu] \int \dfrac{d^3q}{16\pi^2}\dfrac{{\Re}^3(\bm{q}-\bm{k})}{\omega_{q-k}}\dfrac{{\Re}^3(\bm{q})}{\omega_q}\\
&=e^2 tr[\gamma^\mu\gamma^\rho\gamma^\nu\gamma^\lambda] \Pi_{(\rho\lambda)}(k) + m^2e^2 tr[\gamma^\mu\gamma^\nu]\Pi(k) 
\end{align} 

\subsection{Electron self-energy }
The Feynman integral(satisfied Lorenz gauge) of Figure 6(b).
\begin{align}
-i\Sigma(p\mkern-8mu/) = (-ie)^2\int \dfrac{d^4k}{{(2\pi)}^4}\dfrac{-ig_{\mu\nu}{\Re}^3(\bm{k})}{k^2-m^2_\gamma+i\epsilon}\gamma^\mu\dfrac{i{\Re}^3(\bm{p}-\bm{k})}{{p\mkern-8mu/}-{k\mkern-8mu/} - m+i\epsilon}\gamma^\nu
\end{align} 
where $m_\gamma$  is the mass of photon.Above equation can be changed to 
\begin{align}
-i\Sigma(p\mkern-8mu/) = e^2\int \dfrac{d^3k}{16\pi^2}\dfrac{-ig_{\mu\nu}{\Re}^3(\bm{k})}{\omega_k}\gamma^\mu\dfrac{i({p\mkern-8mu/}-{k\mkern-8mu/}+m){\Re}^3(\bm{p}-\bm{k})}{\omega_{p-k}}\gamma^\nu
\end{align} 
with
\begin{align}
\omega_{p-k} &=\sqrt{(\bm{p}-\bm{k})^2+m^2}  \\
\omega_k &=\sqrt{\bm{k}^2} 
\end{align} 

\begin{figure}[htbp]
\centering
\subfigure[Vacuum polarization loop diagram]{
\begin{minipage}[t]{0.5\linewidth}
\centering
\includegraphics[width=1\textwidth]{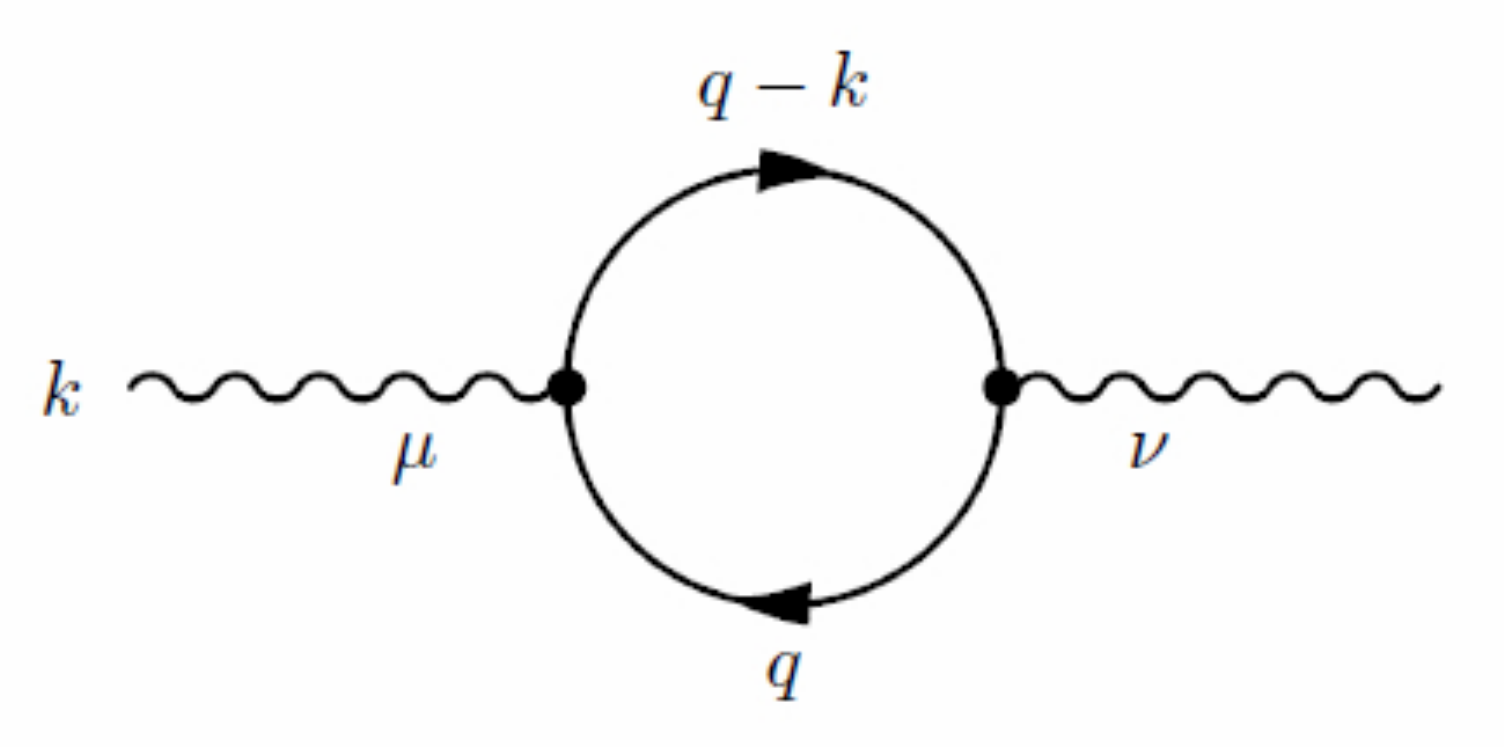}
\end{minipage}%
}%
\subfigure[Electron self-energy loop diagram]{
\begin{minipage}[t]{0.5\linewidth}
\centering
\includegraphics[width=1\textwidth]{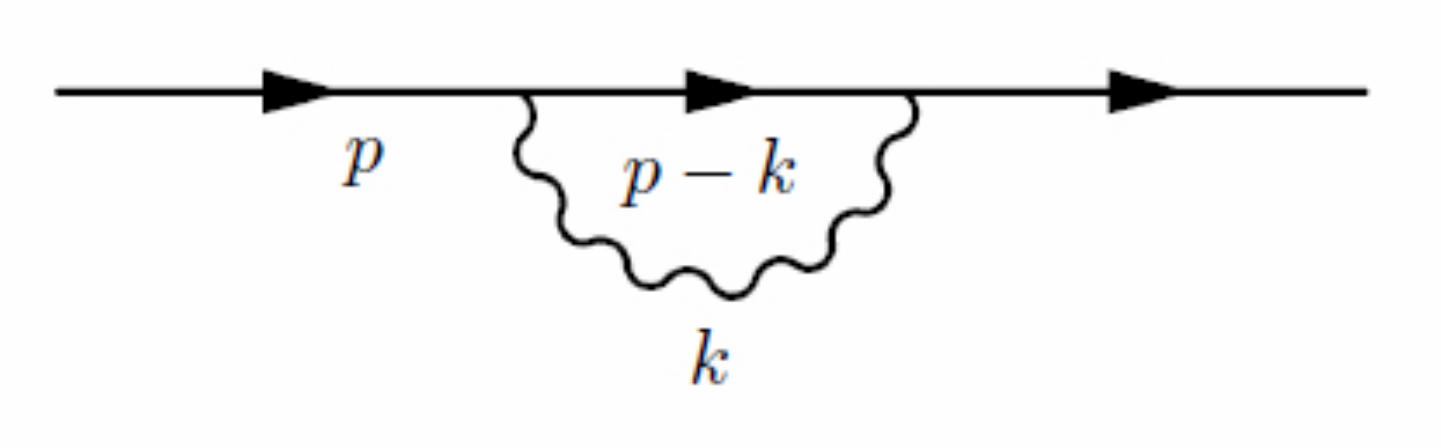}
\end{minipage}%
}%
\centering
\caption{ Loop Diagram}
\end{figure}
\section{Symmetry}
\subsection{The Klein-Gordon field \& the Dirac field}
For equation (3.4), we have
\begin{align}
(\Box  + m^2 )\phi_P= 0
\end{align} 
So $\phi_P$ is satisfied the Klein-Gordon equation.The Klein-Gordon equation
is Lorentz covariance. For equation (3.15), we have
\begin{align}
(i\gamma^\mu\partial_\mu  - m )\psi_P= 0
\end{align} 
and $\psi_P$ is satisfied the Dirac equation.

\subsection{Gauge invariance}
The Lagrangian of scalar field interacting with abelian gauge field
\begin{align}
\mathcal{L} = (D_\mu\phi_P)^\dagger (D^\mu\phi_P) - \dfrac{1}{4}[F_{\mu\nu}]_P[F^{\mu\nu}]_P - V(\phi_P)
\end{align} 
with 
\begin{align}
D^\mu\phi_P  &= (\partial^\mu - ig[A^\mu]_P)\phi_P\\
[F_{\mu\nu}]_P &=[\partial_\mu A_\nu - \partial_\nu A_\mu]_P
\end{align}
where $V(\phi_P)$ have no derivative of $\phi_P$.
Above Lagrangian is invariance under below transformations
\begin{align}
\phi_P \to \phi^\prime_P &= e^{-ig\theta(x)}\phi_P \\
[A^\mu]_P \to [A^{\prime\mu}]_P &=[A^\mu]_P-\partial^\mu\theta
\end{align} 
From the description of this section of  quantization and propagator, the renormalization function  involve  in creation and annihilation operators. Thus above results is obvious. For non-abelian gauge field, it can not use canonical quantization. This paper will not discuss.
\section{Conclusion}
In the field of quantum mechanics, the wave function of a particle with definite position is a Dirac function. Because of the limitation of Planck scale, Dirac function needs to be replaced by $\delta_P$ function.  It is also used in quantum field theory. 

The renormalization function  involve  in creation and annihilation operators, and it is depend on momentum. Under the renormalization transformations, field $\phi_P$ and field $\psi_P$ satisfy the Klein-Gordon equation and the Dirac equation respectively, and the Lagrangian of scalar field interacting with abelian gauge field  is gauge invariant.

For free fields, the new Feynman propagators which product a factor of renormalization function(or combination of renormalization function) are convergent in large momentum. For  interaction fields, it is simple to prove that the  propagators are convergent.  The problem of infinities is the result of introducing  the Dirac delta function in position representation. When we introduce the Dirac delta function in position representation, this will cause momentum probability amplitude  same in any values, and lead to divergence difficulties of large momentum. The introduction of renormalization function can eliminate the ultraviolet divergence, because the limit of  large momentum probability amplitude is  convergent, and $\omega_k$ is bounded for large momentum(detail see Appendix), so new Feynman propagators are convergent.

\appendix
\section{Proof of convergence of  loop diagram}
Proposition\cite{math01}  (Abel-Dirichlet test for convergence of an integral).Suppose that $g$ is monotonic.
Then a sufficient condition for convergence of the improper integral
\begin{align}
\int_a^{+\infty} f(x) g(x)dx
\end{align}
is that the one of the following  conditions hold:
\paragraph{a.}
   (Abel test)the integral $\int_a^{+\infty} f(x) dx$ converges, and the function $g(x)$ is bounded on $[a,+\infty)$.
\paragraph{b.}
  (Dirichlet test)the function  $F(A) = \int_a^{A} f(x) dx$  is bounded on $[a,+\infty)$, the function $g(x)$ tends to zero as $x \to +\infty$, $x {\in}[a,+\infty)$. \\\\
Theorem\cite{Arhipov2006}(Comparison test) $\forall x \in [a,+\infty)$ if $\left|f(x)\right| \leq g(x)$ and integral $\int_a^{+\infty} g(x)dx$ is convergent, then integral $\int_a^{+\infty} f(x)dx$ is convergent.

\subsection{Rectangular coordinate system}
\subsubsection{Vacuum polarization loop diagram}
We will prove that $\Pi_{(\rho\lambda)}(k)$ and $\Pi(k)$ in equation (4.8) are convergent.\\
Assume  $q_c >\left|k^i\right|>0$ with $\bm{k}=k^i=-k_i=(k_x,k_y,k_z)$. So $g(\bm{q}) = \dfrac{1}{\omega_{q-k}\omega_{q}}$  is monotonic on $[q_c,+\infty)$ and $g(\bm{q}) <\dfrac{1}{m^2}$ .
 By comparison test   integral 
\begin{align}
\int_{q_c}^{+\infty}[{\sin\dfrac{l_P(q^i-k^i)}{2\hbar }}/{\dfrac{l_P(q^i-k^i)}{2\hbar}}][{\sin\dfrac{l_Pq^i}{2\hbar }}/{\dfrac{l_Pq^i}{2\hbar}}] dq^i
\end{align}
is convergent.  By Abel test, 
integral 
\begin{align}
\int^{+\infty}_{q_c} \dfrac{d^3q}{16\pi^2}\dfrac{{\Re}^3_{\bot}(\bm{q}-\bm{k})}{\omega_{q-k}}\dfrac{{\Re}^3_{\bot}(\bm{q})}{\omega_q}
\end{align} 
is convergent too.
Similarly we can prove that integral 
\begin{align}
\int_{-\infty}^{q_d} \dfrac{d^3q}{16\pi^2}\dfrac{{\Re}^3_{\bot}(\bm{q}-\bm{k})}{\omega_{q-k}}\dfrac{{\Re}^3_{\bot}(\bm{q})}{\omega_q}
\end{align} 
 is convergent, with $q_d <-\left|k^i\right|<0$.

 Integral 
\begin{align}
\int_{q_c}^{q_d} \dfrac{d^3q}{16\pi^2}\dfrac{{\Re}^3_{\bot}(\bm{q}-\bm{k})}{\omega_{q-k}}\dfrac{{\Re}^3_{\bot}(\bm{q})}{\omega_q}
\end{align} 
is not an improper integral, obviousely it is convergent. Above all integral
\begin{align}
\Pi(k) =  \int \dfrac{d^3q}{16\pi^2}\dfrac{{\Re}^3_{\bot}(\bm{q}-\bm{k})}{\omega_{q-k}}\dfrac{{\Re}^3_{\bot}(\bm{q})}{\omega_q}
\end{align}
is convergent.
From above proof, integrals
\begin{align}
\int_{q_c}^{+\infty}[{\sin\dfrac{l_P(q^i-k^i)}{2\hbar }}/{\dfrac{l_P(q^i-k^i)}{2\hbar}}][{\sin\dfrac{l_Pq^i}{2\hbar }}/{\dfrac{l_Pq^i}{2\hbar}}] dq^i \\
\int^{q_d}_{-\infty}[{\sin\dfrac{l_P(q^i-k^i)}{2\hbar }}/{\dfrac{l_P(q^i-k^i)}{2\hbar}}][{\sin\dfrac{l_Pq^i}{2\hbar }}/{\dfrac{l_Pq^i}{2\hbar}}] dq^i
\end{align}
are convergent.  Functions
\begin{align}
g(q_i-k_i,q_j) &=\sqrt{\dfrac{(q_i-k_i)^2}{(\bm{q}-\bm{k})^2+m^2 }\dfrac{q_j^2}{\bm{q}^2+m^2 }}  \\
g(q_0-k_0,q_j) &=\sqrt{\dfrac{q_j^2}{\bm{q}^2+m^2 }}  \\
g(q_i-k_i,q_0) &=\sqrt{\dfrac{(q_i-k_i)^2}{(\bm{q}-\bm{k})^2+m^2 }}
\end{align} 
are monotonic on $[q_c,+\infty)$ where $q_c>\left|k_i\right|>0$,  and $g(q_i-k_i,q_j)<1$. By Abel test , integral 
\begin{align}
\int_{q_c}^{+\infty} \dfrac{d^3q}{16\pi^2} \dfrac{(q_\rho-k_\rho)q_\lambda{\Re}^3_{\bot}(\bm{q}-\bm{k})}{\omega_{q-k}}\dfrac{{\Re}^3_{\bot}(\bm{q})}{\omega_q}
\end{align}
is convergent. Similarly, integral 
\begin{align}
 \int^{q_d}_{-\infty} \dfrac{d^3q}{16\pi^2} \dfrac{(q_\rho-k_\rho)q_\lambda{\Re}^3_{\bot}(\bm{q}-\bm{k})}{\omega_{q-k}}\dfrac{{\Re}^3_{\bot}(\bm{q})}{\omega_q}
\end{align}
is convergent.
Proper integral  
\begin{align}
\int_{q_c}^{q_d} \dfrac{d^3q}{16\pi^2} \dfrac{(q_\rho-k_\rho)q_\lambda{\Re}^3_{\bot}(\bm{q}-\bm{k})}{\omega_{q-k}}\dfrac{{\Re}^3_{\bot}(\bm{q})}{\omega_q}
\end{align} 
 is convergent.
Above all, improper integral
\begin{align}
 \Pi_{(\rho\lambda)}(k) = \int_{-\infty}^{+\infty} \dfrac{d^3q}{16\pi^2} \dfrac{(q_\rho-k_\rho)q_\lambda{\Re}^3_{\bot}(\bm{q}-\bm{k})}{\omega_{q-k}}\dfrac{{\Re}^3_{\bot}(\bm{q})}{\omega_q}
\end{align}
is convergent. So far, $\Pi^{\mu\nu}(k)$ is convergent.


\begin{figure}[htbp]
\centering
\subfigure[N vertices Feynman loop diagram]{
\begin{minipage}[t]{0.4\linewidth}
\centering
\includegraphics[width=1\textwidth]{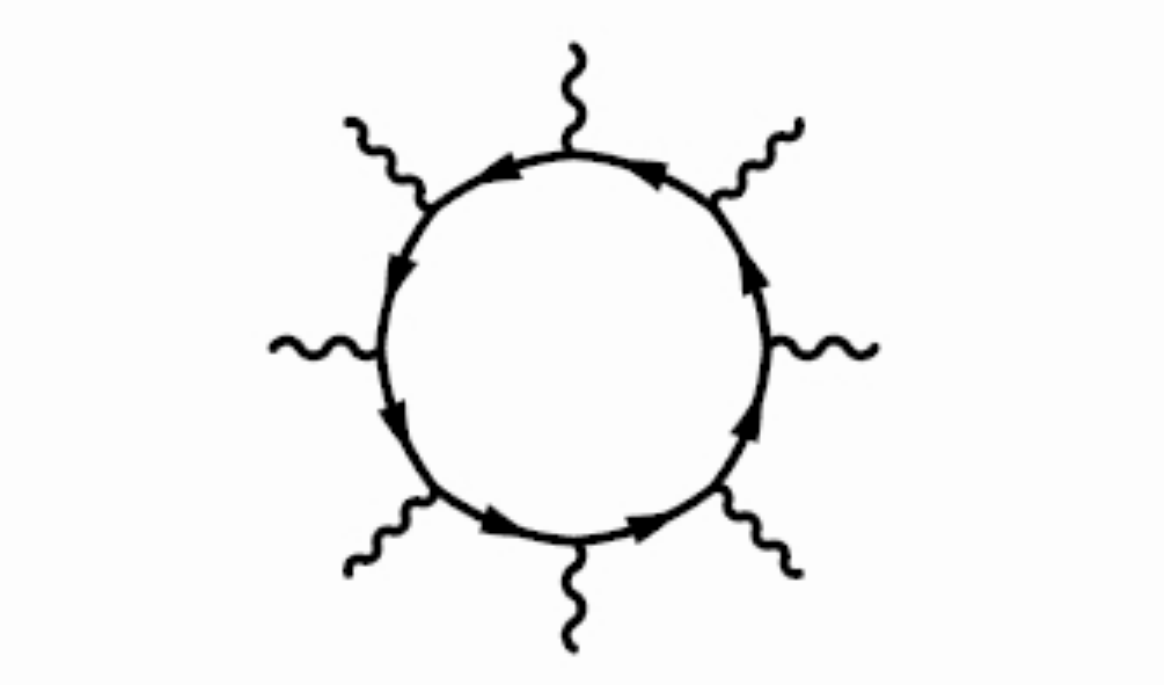}
\end{minipage}%
}%
\subfigure[Baryons mesons interaction loop diagram]{
\begin{minipage}[t]{0.5\linewidth}
\centering
\includegraphics[width=1\textwidth]{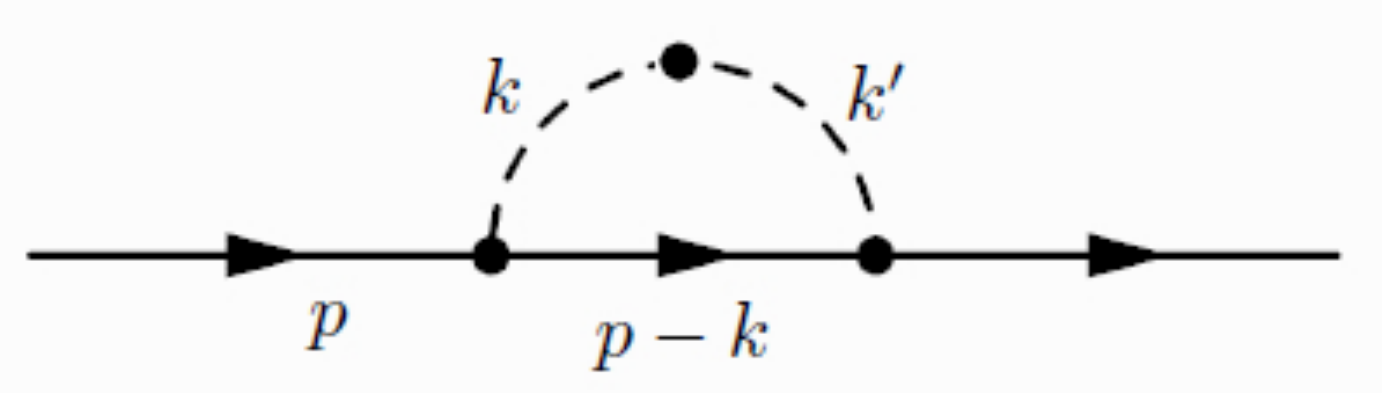}
\end{minipage}%
}%
\centering
\caption{ Feynman Diagram}
\end{figure}
\subsubsection{Fermion loop diagram of  N photon external lines}
 The  Feynman integral of Figure 7 (a) (below equation constant factor omitted. We only prove the case which external photon lines are injection momentum , other cases are similar)
\begin{align}
J_1^{\mu_1\mu_2...\mu_n} &= \int \dfrac{d^4p}{(2\pi)^4}(-1) tr \{ \dfrac{i {\Re}^3_{\bot}(\bm{p})}{ {p\mkern-8mu/}- m + i\epsilon} \gamma^{\mu_1} \dfrac{i {\Re}^3_{\bot}(\bm{p}-\bm{k}_1)}{ {p\mkern-8mu/}-{k\mkern-8mu/}_1- m+ i\epsilon}\gamma^{\mu_2}... \nonumber\\
&\times  \dfrac{i {\Re}^3_{\bot}(\bm{p}-\bm{k}_1-...-\bm{k}_{n-1})}{ {p\mkern-8mu/}-{k\mkern-8mu/}_1-...-{k\mkern-8mu/}_{n-1}- m+ i\epsilon} \} \delta^4(k_1+ k_2+...+ k_n)\\
 &= \int \dfrac{d^3p}{(2\pi)^4}(-1)(\pi)^n tr \{  ( {p\mkern-8mu/}+ m) \gamma^{\mu_1} ({p\mkern-8mu/}-{k\mkern-8mu/}_1+ m)\gamma^{\mu_2}... ({p\mkern-8mu/}-{k\mkern-8mu/}_1-...-{k\mkern-8mu/}_{n-1}+m) \}   \nonumber\\
&\times  \dfrac{{\Re}^3_{\bot}(\bm{p})}{\omega_p}  \dfrac{{\Re}^3_{\bot}(\bm{p}-\bm{k}_1)}{\omega_{p-k_1}}... \dfrac{{\Re}^3_{\bot}(\bm{p}-\bm{k}_1-...-\bm{k}_{n-1})}{\omega_{p-k_1-...-k_{n-1}}} \delta^4(k_1+ k_2+...+ k_n)
\end{align}
with
\begin{align}
\omega_p &=\sqrt{\bm{p}^2+ m^2} \\
\omega_{p-k_1} &=\sqrt{(\bm{p}-\bm{k}_1)^2+m^2}  \\
...\nonumber
\end{align} 
If we can prove integral
\begin{align}
 &\int \dfrac{d^3p}{(2\pi)^3}
 \dfrac{{\Re}^3_{\bot}(\bm{p})}{\omega_p}  \dfrac{{\Re}^3_{\bot}(\bm{p}-\bm{k}_1)}{\omega_{p-k_1}}... \dfrac{{\Re}^3_{\bot}(\bm{p}-\bm{k}_1-...-\bm{k}_{n-1})}{\omega_{p-k_1-...-k_{n-1}}} \nonumber\\
 &\times p_{\nu_1}(p_{\nu_2}-k_{1{\nu_2}})...(p_{\nu_n}-k_{1{\nu_n}}...-k_{(n-1)\nu_n})
\end{align} 
is convergent, then equation (A.16) is convergent.
Function
\begin{align}
g(p_i,p_{i_1}...,p_{i_{n-1}}) =\sqrt{\dfrac{p_i^2}{\bm{p}^2+m^2 } \dfrac{(p_{i_1}-k_{1{i_1}})^2}{(\bm{p}-\bm{k_1})^2+m^2 }...\dfrac{(p_{i_{n-1}}-k_{1{i_{n-1}}}-...-k_{(n-1){i_{n-1}}})^2}{(\bm{p}-\bm{k}_1-...-\bm{k}_{n-1})^2+m^2 }}
\end{align}
is monotonic on $p_i \in [q_c,+\infty)$ where $q_c >\left|k_{1i}\right|+\left|k_{2i}\right|+...+\left|k_{(n-1)i}\right|>0$, and $g(p_i,p_{i_1}...,p_{i_{n-1}})<1$. Similar handle procedure with (A.7) (A.8) , we can obtain integrals 
\begin{align}
&\int_{q_c}^{+\infty}[{\sin\dfrac{l_Pp^i}{2\hbar }}/{\dfrac{l_Pp^i}{2\hbar}}] [{\sin\dfrac{l_P(p^i-k^i_{1})}{2\hbar }}/{\dfrac{l_P(p^i-k^i_1)}{2\hbar}}]...\nonumber\\&\times [{\sin\dfrac{l_P(p^i-k^i_1-...-k^i_{n-1})}{2\hbar }}/{\dfrac{l_P(p^i-k^i_1-...-k^i_{n-1})}{2\hbar}}]dp^i \\
&\int^{q_d}_{-\infty}[{\sin\dfrac{l_Pp^i}{2\hbar }}/{\dfrac{l_Pp^i}{2\hbar}}] [{\sin\dfrac{l_P(p^i-k^i_1)}{2\hbar }}/{\dfrac{l_P(p^i-k^i_1)}{2\hbar}}]...\nonumber\\&\times [{\sin\dfrac{l_P(p^i-k^i_1-...-k^i_{n-1})}{2\hbar }}/{\dfrac{l_P(p^i-k^i_1-...-k^i_{n-1})}{2\hbar}}]dp^i 
\end{align}
are convergent(where $q_d <-\left|k^i_1\right|-\left|k^i_2\right|-...-\left|k^i_{n-1}\right|<0$). By Abel test, integral
\begin{align}
 &\int_{q_c}^{+\infty} \dfrac{d^3p}{(2\pi)^3}
 \dfrac{{\Re}^3_{\bot}(\bm{p})}{\omega_p}  \dfrac{{\Re}^3_{\bot}(\bm{p}-\bm{k}_1)}{\omega_{p-k_1}}... \dfrac{{\Re}^3_{\bot}(\bm{p}-\bm{k}_1-...-\bm{k}_{n-1})}{\omega_{p-k_1-...-k_{n-1}}} \nonumber\\
 &\times p_{\nu_1}(p_{\nu_2}-k_{1{\nu_2}})...(p_{\nu_n}-k_{1{\nu_n}}...-k_{(n-1)\nu_n})
\end{align} 
is convergent. Similarly, integral
\begin{align}
 &\int^{q_d}_{-\infty} \dfrac{d^3p}{(2\pi)^3}
 \dfrac{{\Re}^3_{\bot}(\bm{p})}{\omega_p}  \dfrac{{\Re}^3_{\bot}(\bm{p}-\bm{k}_1)}{\omega_{p-k_1}}... \dfrac{{\Re}^3_{\bot}(\bm{p}-\bm{k}_1-...-\bm{k}_{n-1})}{\omega_{p-k_1-...-k_{n-1}}} \nonumber\\
 &\times p_{\nu_1}(p_{\nu_2}-k_{1{\nu_2}})...(p_{\nu_n}-k_{1{\nu_n}}...-k_{(n-1)\nu_n})
\end{align}
convergent. We obtain integral (A.20) is convergent. Thus equation (A.16) is convergent.
\subsubsection{Electron self-energy  loop diagram}
From equation (4.10), we have
\begin{align}
-i\Sigma(p\mkern-8mu/) &=e^2g_{\mu\nu}\int \dfrac{d^3k}{16\pi^2}\dfrac{{\Re}^3_{\bot}(\bm{k})}{\omega_k}\gamma^\mu\dfrac{({p\mkern-8mu/}-{k\mkern-8mu/} + m){\Re}^3_{\bot}(\bm{p}-\bm{k})}{\omega_{p-k}}\gamma^\nu \\
&= e^2g_{\mu\nu}\gamma^\mu\gamma^\rho\gamma^\nu\int \dfrac{d^3k}{16\pi^2}\dfrac{{\Re}^3_{\bot}(\bm{k})}{\omega_k}\dfrac{(p_\rho-k_\rho){\Re}^3_{\bot}(\bm{p}-\bm{k})}{\omega_{p-k}} \\
&+ e^2 m g_{\mu\nu}\gamma^\mu\gamma^\nu\int \dfrac{d^3k}{16\pi^2}\dfrac{{\Re}^3_{\bot}(\bm{k})}{\omega_k}\dfrac{{\Re}^3_{\bot}(\bm{p}-\bm{k})}{\omega_{p-k}} 
\end{align} 
Compare  (A.27) (A.28) with (4.6)(4.7) and from above proof, we can obtain $-i\Sigma(p\mkern-8mu/)$ is convergent on large momentum.
\subsubsection{Arbitrary order case}
Consider integral
\begin{align}
 &\int {d^3p}
 \dfrac{{\Re}^3_{\bot}(\bm{p}-\bm{k}_1)}{\omega_{p - k_1}}  \dfrac{{\Re}^3_{\bot}(\bm{p}-\bm{k}_2)}{\omega_{p-k_2}}... \dfrac{{\Re}^3_{\bot}(\bm{p}-\bm{k}_{n})}{\omega_{p-k_{n}}} \nonumber\\
 &\times (p_{\nu_1}-k_{1{\nu_1}})(p_{\nu_2}-k_{2{\nu_2}})...(p_{\nu_n}-k_{n{\nu_n}})
\end{align} 
with
\begin{align}
\omega_{p-k_j} &=\sqrt{(\bm{p}-\bm{k}_j)^2+m_j^2}  \\
j &=1,2,...,n\nonumber
\end{align} 
From proof procedure of (A.20), we obtain above integral is convergent. 
\subsection{Cylindrical coordinate system}
\subsubsection{Vacuum polarization loop diagram}
Firstly we may proove that integral
\begin{align}
\int_{-\infty}^{+\infty}\int_{-\infty}^{+\infty}\int_{-\infty}^{+\infty} \sideset{_0}{}{\mathop{F_1}}(2;-\frac{l_P^2(p_x^2+p_y^2)}{16\hbar^2})[{\sin\dfrac{l_Pp_z}{2\hbar }}/{\dfrac{l_Pp_z}{2\hbar}}] dp_xdp_ydp_z
\end{align} 
is convergent. From Cauchy-Hadamard  theorem\cite{Arhipov2006}, we obtain power series $ \sideset{_0}{}{\mathop{F_1}}(2;-\frac{l_P^2(p_x^2+p_y^2)}{16\hbar^2})$ is convergent for any $p_x$ and $p_y$. 
By calculate 
\begin{align}
&\int_{-\infty}^{+\infty}\int_{-\infty}^{+\infty} \sideset{_0}{}{\mathop{F_1}}(2;-\frac{l_P^2(p_x^2+p_y^2)}{16\hbar^2}) dp_xdp_y \nonumber\\
&= \int_0^{+\infty}\int_0^{2\pi} r\sum_{n=0}^\infty \frac{(-1)^n}{n!(n+1)!}(\frac{l_P^2 r^2}{16\hbar^2})^n drd\theta \\
&=  \frac{16\pi\hbar^2}{l_P^2}\lim_{r \to +\infty}\sum_{n=0}^\infty \frac{(-1)^n}{[(n+1)!]^2}r^{n+1} \\
&=  \frac{16\pi\hbar^2}{l_P^2} \nonumber
\end{align}
And integral
\begin{align}
\int_{-\infty}^{+\infty}{\sin\dfrac{l_Pp_z}{2\hbar }}/{\dfrac{l_Pp_z}{2\hbar}} dp_z
\end{align} 
is convergent. Thus (A.31) is convergent. \\
We can prove that integral
\begin{align}
\int^{+\infty}_{q_c} \dfrac{d^3q}{16\pi^2}\dfrac{{\Re}^3_{Cylin}(\bm{q}-\bm{k})}{\omega_{q-k}}\dfrac{{\Re}^3_{Cylin}(\bm{q})}{\omega_q}
\end{align} 
is convergent. As power series $\sideset{_0}{}{\mathop{F_1}}(2;-\frac{l_P^2r^2}{16\hbar^2})$ is convergent for any $r\in Real$, and 
\begin{align}
\lim_{r \to +\infty}\sideset{_0}{}{\mathop{F_1}}(2;-\frac{l_P^2r^2}{16\hbar^2}) = 0
\end{align} 
Assume $\left|\sideset{_0}{}{\mathop{F_1}}(2;-\frac{l_P^2r^2}{16\hbar^2})\right| \leq  F_{max}$. From Cauchy-Hadamard  theorem, obtain integral
\begin{align}
\int^{+\infty}_{q_c} d^3q {\Re}^3_{Cylin}(\bm{q}-\bm{k}) {\Re}^3_{Cylin}(\bm{q})
\end{align} 
is convergent. By Abel test, integral (A.35) is convergent.\\
The remain steps refer to similar steps in rectangular coordinate system, we can obtain $\Pi^{\mu\nu}(k)$ is convergent in cylindrical coordinate system.
\subsubsection{Electron self-energy  loop diagram}
The process of proof similar with case in  the previous section which use Cauchy-Hadamard  theorem and Abel-Dirichlet test.

\subsubsection{Arbitrary order case}
Consider integral
\begin{align}
 &\int {d^3p}
 \dfrac{{\Re}^3_{Cylin}(\bm{p}-\bm{k}_1)}{\omega_{p - k_1}}  \dfrac{{\Re}^3_{Cylin}(\bm{p}-\bm{k}_2)}{\omega_{p-k_2}}... \dfrac{{\Re}^3_{Cylin}(\bm{p}-\bm{k}_{n})}{\omega_{p-k_{n}}} \nonumber\\
 &\times (p_{\nu_1}-k_{1{\nu_1}})(p_{\nu_2}-k_{2{\nu_2}})...(p_{\nu_n}-k_{n{\nu_n}})
\end{align} 
with
\begin{align}
\omega_{p-k_j} &=\sqrt{(\bm{p}-\bm{k}_j)^2+m_j^2}  \\
j &=1,2,...,n\nonumber
\end{align} 
By  Cauchy-Hadamard  theorem, integral
\begin{align}
 &\int {d^3p}
 {\Re}^3_{Cylin}(\bm{p}-\bm{k}_1)  {\Re}^3_{Cylin}(\bm{p}-\bm{k}_2)... {\Re}^3_{Cylin}(\bm{p}-\bm{k}_{n})
\end{align} 
is convergent. Again by Abel-Dirichlet test to prove that integral (A.38) is convergent.
\section{Calculate the masses of octet baryons from chiral perturbation theory}
\subsection{Rectangular coordinate system}
In this section,the authors will calculate one loop graph from renormalization function  ${\Re}^3_{\bot}(\textbf{\textit{k}})$, and calculate the masses of octet baryons from chiral perturbation theory\cite{ref13,ref14}.
The Figure 7 (b)  loop graph contribution
\begin{align}
\sigma[M_1,M_2]  &= -\gamma^\mu \gamma_5 \dfrac{1 }{16\pi} \int {d^3k}(\overline{m}+{p\mkern-8mu/}-{k\mkern-8mu/})(k_\mu-\Delta_\mu)k_\nu\nonumber\\
 &\times \dfrac{{\Re}^3_{\bot}(\textbf{\textit{p}}-\textbf{\textit{k}}){\Re}^3_{\bot}(\textbf{\textit{k}}){\Re}^3_{\bot}(\textbf{\textit{k}}-\bm{\Delta}) }{\omega_{p-k}\omega_{k}\omega_{k-\Delta}} \gamma^\nu \gamma_5
\end{align} 
with
\begin{align}
\omega_{p-k}&=\sqrt{(\textbf{\textit{p}}-\textbf{\textit{k}})^2+\overline{m}^2} \\
\omega_k&=\sqrt{\textbf{\textit{k}}^2+M^2_1} \\
\omega_{k-\Delta}&=\sqrt{(\textbf{\textit{k}}-\bm{\Delta})^2+M^2_2} 
\end{align}
Octet baryons mean mass
\begin{align}
\overline{m} = 1151 MeV
\end{align}
Pseudoscalar mesons experimental values of masses
\begin{align}
M_{\pi^0} &= 134.96 MeV, M_{\eta}=548.8 MeV,  \nonumber\\
M_{\pi^+} &= M_{\pi^-}=139.57 MeV,  M_{K^+}= M_{K^-}= 493.71 MeV  \nonumber\\
 M_{K^0} &= M_{\overline{K}^0}= 497.70 MeV  \nonumber
\end{align}
The calculation results for loop graph
\begin{align}
\sigma[M_\eta,M_\eta]  &= -577.2830,  \sigma[M_{\pi^0},M_\eta]  = -574.3426,  \nonumber\\
 \sigma[M_{\pi^0},M_{\pi^0}]  &= -565.3559,  \sigma[M_{\pi^+},M_{\pi^-}]  = -566.1214,  \nonumber\\
 \sigma[M_{K^+},M_{K^-}]  &= -577.0600,  \sigma[M_{K^0},M_{\overline{K}^0}]  = -577.0778   
\end{align} 
Quark experimental values of masses
\begin{align}
m_u = 2 MeV, m_d = 5 MeV, m_s = 150 MeV  
\end{align} 
Constants 
\begin{align}
g_A = 1.25, B =2.1 MeV, f_\pi = 93.3 MeV
\end{align} 
Here assume constant $B$ absorb the constant from calculating loop graph contribution $\sigma[M_1,M_2 ]$ (if $B$ keep unchanged, then need add a new constant $g_{RF}$, satisfied $B*g_{RF} =2.1$, thus keep calculation results unchanged).
Set 
\begin{align}
c_r = 547.8883,d=0.6983
\end{align} 
Which make below equation
\begin{align}
\Delta m_{RF} = \sum_{B=1}^8 Abs(m^{cal}_B - m^{exp}_B )
\end{align} 
a  minimum value ($\Delta m_{RF}=58.78$).With $m^{exp}_B$ represent the experimental values of  oectet baryons masses, and $m^{cal}_B$ represent the theoretical calculation values. From above parameter setting, we can obtain the theoretical calculation results(the calculation formulas of mass splitting refer to citation\cite{ref13})

\begin{align}
m_p  &=938.28 MeV, m_n=940.31 MeV,  \nonumber\\
m_\Lambda &=1092.02MeV,m_{\Sigma^0}  =1199.09MeV, \nonumber\\
m_{\Sigma^+} &=1178.02MeV,m_{\Sigma^-} =1183.62MeV,\nonumber\\
m_{\Xi^0}  &=1317.72MeV, m_{\Xi^-} =1321.30MeV 
\end{align} 

\subsection{Cylindrical coordinate system}
In this section, the authors will calculate the results  in cylindrical coordinate system. 
The Figure 7 (b)  loop graph contribution
\begin{align}
\sigma[M_1,M_2]  &= -\gamma^\mu \gamma_5 \dfrac{1 }{16\pi} \int {d^3k}(\overline{m}+{p\mkern-8mu/}-{k\mkern-8mu/})(k_\mu-\Delta_\mu)k_\nu\nonumber\\
 &\times \dfrac{{\Re}^3_{cylin}(\textbf{\textit{p}}-\textbf{\textit{k}}){\Re}^3_{cylin}(\textbf{\textit{k}}){\Re}^3_{cylin}(\textbf{\textit{k}}-\bm{\Delta}) }{\omega_{p-k}\omega_{k}\omega_{k-\Delta}} \gamma^\nu \gamma_5
\end{align} 
By integrate highly ocillatory functions, we obtain calculation results for loop graph
\begin{align}
\sigma[M_\eta,M_\eta]  &= -765.8240,  \sigma[M_{\pi^0},M_\eta]  = -760.9211,  \nonumber\\
 \sigma[M_{\pi^0},M_{\pi^0}]  &= -745.9441,  \sigma[M_{\pi^+},M_{\pi^-}]  = -747.2172,  \nonumber\\
 \sigma[M_{K^+},M_{K^-}]  &= -765.4512,  \sigma[M_{K^0},M_{\overline{K}^0}]  = -765.4819  
\end{align} 
By similar methods, set
\begin{align}
c_r =  0.7156,d = 0.6862, B = 0.12
\end{align} 
and obtain 
\begin{align}
\Delta m_{RF} = 59.565
\end{align} 
it is a minimum value. Input these parameters to the calculation formulas of mass splitting
\begin{align}
m_p  &=938.02 MeV, m_n=939.93 MeV,  \nonumber\\
m_\Lambda &=1086.98MeV,m_{\Sigma^0}  =1208.07MeV, \nonumber\\
m_{\Sigma^+} &=1184.57MeV,m_{\Sigma^-} =1190.12MeV,\nonumber\\
m_{\Xi^0}  &=1317.65MeV, m_{\Xi^-} =1321.30MeV 
\end{align} 
\subsection{The calculation formulas of mass splitting}
Here we list the calculation formulas of mass splitting, detail see\cite{ref13}
\begin{align}
\Delta m_p&=  \dfrac{B\overline{m}}{f^2_\pi}\{c_3[-\dfrac{c_r \kappa_2}{3}-\dfrac{g^2_A\kappa_2}{4}\sigma[M_{K^0},M_{\overline{K}^0}]+(\dfrac{d^2}{3}-\dfrac{2dg_A}{3}+\dfrac{g^2_A}{4}) \sqrt{\kappa_4}\sigma[M_{\pi^0},M_\eta] \nonumber\\
&+(\dfrac{2d^2}{3}-dg_A+\dfrac{g_A^2}{2})\kappa_2\sigma[M_{K^+},M_{K^-}]]+c_8[-\dfrac{5c_r\kappa_2}{3\sqrt{3}}-\dfrac{2c_r\kappa_4}{3\sqrt{3}}+(\dfrac{d^2}{2\sqrt{3}}-\dfrac{dg_A}{2\sqrt{3}} \nonumber\\
&+\dfrac{g^2_A}{8\sqrt{3}})\sigma[M_{\pi^0},M_{\pi^0}]+(-\dfrac{d^2}{6\sqrt{3}}+\dfrac{dg_A}{2\sqrt{3}}-\dfrac{\sqrt{3}g^2_A}{8})\kappa_4\sigma[M_\eta,M_\eta]-\dfrac{g^2_A}{4\sqrt{3}}\kappa_2\sigma[M_{K^0},M_{\overline{K}^0}] \nonumber\\
&+(-\dfrac{2d^2}{3\sqrt{3}}+\dfrac{dg_A}{\sqrt{3}}-\dfrac{g^2_A}{2\sqrt{3}})\kappa_2\sigma[M_{K^+},M_{K^-}]+(\dfrac{2d^2}{\sqrt{3}}-\dfrac{2dg_A}{\sqrt{3}}+\dfrac{g^2_A}{2\sqrt{3}})\sigma[M_{\pi^+},M_{\pi^-}] ]\} 
\end{align} 

\begin{align}
\Delta m_n&=  \dfrac{B\overline{m}}{f^2_\pi}\{c_3[\dfrac{c_r \kappa_2}{3}-(\dfrac{2d^2}{3}-dg_A+\dfrac{g^2_A}{2}) \kappa_2\sigma[M_{K^0},M_{\overline{K}^0}]-(\dfrac{d^2}{3}-\dfrac{2dg_A}{3}+\dfrac{g^2_A}{4}) \sqrt{\kappa_4} \nonumber\\
&\times\sigma[M_{\pi^0},M_\eta]+\dfrac{g_A^2}{4}\kappa_2\sigma[M_{K^+},M_{K^-}]]+c_8[-\dfrac{5c_r\kappa_2}{3\sqrt{3}}-\dfrac{2c_r\kappa_4}{3\sqrt{3}}+(\dfrac{d^2}{2\sqrt{3}}-\dfrac{dg_A}{2\sqrt{3}} \nonumber\\
&+\dfrac{g^2_A}{8\sqrt{3}})\sigma[M_{\pi^0},M_{\pi^0}]+(-\dfrac{d^2}{6\sqrt{3}}+\dfrac{dg_A}{2\sqrt{3}}-\dfrac{\sqrt{3}g^2_A}{8})\kappa_4\sigma[M_\eta,M_\eta]+(-\dfrac{2d^2}{3\sqrt{3}}+\dfrac{dg_A}{\sqrt{3}}-\dfrac{g^2_A}{2\sqrt{3}}) \nonumber\\
&\times\kappa_2\sigma[M_{K^0},M_{\overline{K}^0}]-\dfrac{g^2_A}{4\sqrt{3}}\kappa_2\sigma[M_{K^+},M_{K^-}]+(\dfrac{2d^2}{\sqrt{3}}-\dfrac{2dg_A}{\sqrt{3}}+\dfrac{g^2_A}{2\sqrt{3}})\sigma[M_{\pi^+},M_{\pi^-}] ]\} 
\end{align} 

\begin{align}
\Delta m_\Lambda&=  \dfrac{B\overline{m}}{f^2_\pi}\{c_3[(-\dfrac{5d^2}{6}+\dfrac{3dg_A}{2}-\dfrac{3g^2_A}{4}) \kappa_2\sigma[M_{K^0},M_{\overline{K}^0}]+(\dfrac{5d^2}{6}-\dfrac{3dg_A}{2}+\dfrac{3g^2_A}{4})\kappa_2  \nonumber\\
&\times\sigma[M_{K^+},M_{K^-}]]+c_8[\dfrac{5c_r}{6\sqrt{3}}-\dfrac{5c_r\kappa_2}{3\sqrt{3}}-\dfrac{c_r\kappa_4}{6\sqrt{3}}+\dfrac{d^2}{6\sqrt{3}}\sigma[M_{\pi^0},M_{\pi^0}]-\dfrac{d^2}{6\sqrt{3}}\kappa_4\sigma[M_\eta,M_\eta] \nonumber\\
&+(-\dfrac{5d^2}{6\sqrt{3}}+\dfrac{\sqrt{3}dg_A}{2}-\dfrac{\sqrt{3}g^2_A}{4})\kappa_2(\sigma[M_{K^0},M_{\overline{K}^0}]+\sigma[M_{K^+},M_{K^-}] ) +\dfrac{2d^2}{3\sqrt{3}}\sigma[M_{\pi^+},M_{\pi^-}] ]\} 
\end{align} 

\begin{align}
\Delta m_{\Sigma^0}&=  \dfrac{B\overline{m}}{f^2_\pi}\{c_3[(-\dfrac{d^2}{2}+\dfrac{dg_A}{2}-\dfrac{g^2_A}{4}) \kappa_2\sigma[M_{K^0},M_{\overline{K}^0}]+(\dfrac{d^2}{2}-\dfrac{dg_A}{2}+\dfrac{g^2_A}{4})\kappa_2  \nonumber\\
&\times\sigma[M_{K^+},M_{K^-}]]+c_8[\dfrac{13c_r}{6\sqrt{3}}-\dfrac{c_r\kappa_2}{\sqrt{3}}-\dfrac{c_r\kappa_4}{6\sqrt{3}}+\dfrac{d^2}{6\sqrt{3}}\sigma[M_{\pi^0},M_{\pi^0}]-\dfrac{d^2}{6\sqrt{3}}\kappa_4\sigma[M_\eta,M_\eta] \nonumber\\
&+(-\dfrac{d^2}{2\sqrt{3}}+\dfrac{dg_A}{2\sqrt{3}}-\dfrac{g^2_A}{4\sqrt{3}})\kappa_2(\sigma[M_{K^0},M_{\overline{K}^0}]+\sigma[M_{K^+},M_{K^-}]) \nonumber\\
&+(\dfrac{2d^2}{\sqrt{3}}-\dfrac{4dg_A}{\sqrt{3}}+\dfrac{2g_A^2}{\sqrt{3}})\sigma[M_{\pi^+},M_{\pi^-}] ]\} 
\end{align} 

\begin{align}
\Delta m_{\Sigma^+}&=  \dfrac{B\overline{m}}{f^2_\pi}\{c_3[-c_r\kappa_2-\dfrac{c_r\sqrt{\kappa_4}}{3}-\dfrac{g^2_A}{4}\kappa_2\sigma[M_{K^0},M_{\overline{K}^0}]+ (\dfrac{d^2}{3}-\dfrac{dg_A}{3})\sqrt{\kappa_4}\sigma[M_{\pi^0},M_\eta]  \nonumber\\
&+(d^2-dg_A+\dfrac{g^2_A}{4})\kappa_2\sigma[M_{K^+},M_{K^-}]]+c_8[\dfrac{11c_r}{6\sqrt{3}}-\dfrac{c_r\kappa_2}{\sqrt{3}}-\dfrac{c_r\kappa_4}{6\sqrt{3}}+(\dfrac{d^2}{2\sqrt{3}}-\dfrac{dg_A}{\sqrt{3}}+\dfrac{g_A^2}{2\sqrt{3}}) \nonumber\\
&\times\sigma[M_{\pi^0},M_{\pi^0}]-\dfrac{d^2}{6\sqrt{3}}\kappa_4\sigma[M_\eta,M_\eta]-\dfrac{g^2_A}{4\sqrt{3}}\kappa_2\sigma[M_{K^0},M_{\overline{K}^0}]+(-\dfrac{d^2}{\sqrt{3}}+\dfrac{dg_A}{\sqrt{3}}-\dfrac{g_A^2}{4\sqrt{3}})\kappa_2\nonumber\\
&\times\sigma[M_{K^+},M_{K^-}]+(\dfrac{4d^2}{3\sqrt{3}}-\dfrac{2dg_A}{\sqrt{3}}+\dfrac{g_A^2}{\sqrt{3}})\sigma[M_{\pi^+},M_{\pi^-}] ]\} 
\end{align} 

\begin{align}
\Delta m_{\Sigma^-}&=  \dfrac{B\overline{m}}{f^2_\pi}\{c_3[c_r\kappa_2+\dfrac{c_r\sqrt{\kappa_4}}{3}+(-d^2+dg_A-\dfrac{g^2_A}{4})\kappa_2\sigma[M_{K^0},M_{\overline{K}^0}]+ (-\dfrac{d^2}{3}+\dfrac{dg_A}{3})\sqrt{\kappa_4}  \nonumber\\
&\times\sigma[M_{\pi^0},M_\eta]+\dfrac{g^2_A}{4}\kappa_2\sigma[M_{K^+},M_{K^-}]]+c_8[\dfrac{11c_r}{6\sqrt{3}}-\dfrac{c_r\kappa_2}{\sqrt{3}}-\dfrac{c_r\kappa_4}{6\sqrt{3}}+(\dfrac{d^2}{2\sqrt{3}}-\dfrac{dg_A}{\sqrt{3}}+\dfrac{g_A^2}{2\sqrt{3}}) \nonumber\\
&\times\sigma[M_{\pi^0},M_{\pi^0}]-\dfrac{d^2}{6\sqrt{3}}\kappa_4\sigma[M_\eta,M_\eta]+(-\dfrac{d^2}{\sqrt{3}}+\dfrac{dg_A}{\sqrt{3}}-\dfrac{g_A^2}{4\sqrt{3}})\kappa_2\sigma[M_{K^0},M_{\overline{K}^0}]-\dfrac{g^2_A}{4\sqrt{3}}\kappa_2\nonumber\\
&\times\sigma[M_{K^+},M_{K^-}]+(\dfrac{4d^2}{3\sqrt{3}}-\dfrac{2dg_A}{\sqrt{3}}+\dfrac{g_A^2}{\sqrt{3}})\sigma[M_{\pi^+},M_{\pi^-}] ]\} 
\end{align} 

\begin{align}
\Delta m_{\Xi^0}&=  \dfrac{B\overline{m}}{f^2_\pi}\{c_3[-\dfrac{2c_r\kappa_2}{3}-\dfrac{c_r\sqrt{\kappa_4}}{3}+(-\dfrac{2d^2}{3}+dg_A-\dfrac{g^2_A}{2})\kappa_2\sigma[M_{K^0},M_{\overline{K}^0}]+ (\dfrac{d^2}{3}-\dfrac{dg_A}{4})\sqrt{\kappa_4}  \nonumber\\
&\times\sigma[M_{\pi^0},M_\eta]+(d^2-dg_A+\dfrac{g^2_A}{4})\kappa_2\sigma[M_{K^+},M_{K^-}]]+c_8[\dfrac{5c_r}{2\sqrt{3}}-\dfrac{2c_r\kappa_2}{3\sqrt{3}}-\dfrac{c_r\kappa_4}{6\sqrt{3}}+\dfrac{g_A^2}{8\sqrt{3}} \nonumber\\
&\times\sigma[M_{\pi^0},M_{\pi^0}]+(-\dfrac{2d^2}{3\sqrt{3}}+\dfrac{dg_A}{\sqrt{3}}-\dfrac{\sqrt{3}g_A^2}{8})\kappa_4\sigma[M_\eta,M_\eta]+(-\dfrac{2d^2}{3\sqrt{3}}+\dfrac{dg_A}{\sqrt{3}}-\dfrac{g_A^2}{2\sqrt{3}})\kappa_2\nonumber\\
&\times\sigma[M_{K^0},M_{\overline{K}^0}]+(-\dfrac{d^2}{\sqrt{3}}+\dfrac{dg_A}{\sqrt{3}}-\dfrac{g_A^2}{4\sqrt{3}})\kappa_2\sigma[M_{K^+},M_{K^-}]+\dfrac{g_A^2}{2\sqrt{3}}\sigma[M_{\pi^+},M_{\pi^-}] ]\} 
\end{align} 

\begin{align}
\Delta m_{\Xi^-}&=  \dfrac{B\overline{m}}{f^2_\pi}\{c_3[\dfrac{2c_r\kappa_2}{3}+\dfrac{c_r\sqrt{\kappa_4}}{3}+(-d^2+dg_A-\dfrac{g^2_A}{4})\kappa_2\sigma[M_{K^0},M_{\overline{K}^0}]+ (-\dfrac{d^2}{3}+\dfrac{dg_A}{4})\sqrt{\kappa_4}  \nonumber\\
&\times\sigma[M_{\pi^0},M_\eta]+(\dfrac{2d^2}{3}-dg_A+\dfrac{g^2_A}{2})\kappa_2\sigma[M_{K^+},M_{K^-}]]+c_8[\dfrac{5c_r}{2\sqrt{3}}-\dfrac{2c_r\kappa_2}{3\sqrt{3}}-\dfrac{c_r\kappa_4}{6\sqrt{3}}+\dfrac{g_A^2}{8\sqrt{3}} \nonumber\\
&\times\sigma[M_{\pi^0},M_{\pi^0}]+(-\dfrac{2d^2}{3\sqrt{3}}+\dfrac{dg_A}{\sqrt{3}}-\dfrac{\sqrt{3}g_A^2}{8})\kappa_4\sigma[M_\eta,M_\eta]+(-\dfrac{d^2}{\sqrt{3}}+\dfrac{dg_A}{\sqrt{3}}-\dfrac{g_A^2}{4\sqrt{3}})\kappa_2\nonumber\\
&\times\sigma[M_{K^0},M_{\overline{K}^0}]+(-\dfrac{2d^2}{3\sqrt{3}}+\dfrac{dg_A}{\sqrt{3}}-\dfrac{g_A^2}{2\sqrt{3}})\kappa_2\sigma[M_{K^+},M_{K^-}]+\dfrac{g_A^2}{2\sqrt{3}}\sigma[M_{\pi^+},M_{\pi^-}] ]\} 
\end{align} 




\bibliographystyle{apsrev4-2}
\bibliography{jhep_manuscipts}


\end{document}